\documentclass[12pt]{article}
\usepackage{amsmath,amssymb,amsthm}
\usepackage{graphicx,subfigure}
\usepackage[multiple]{footmisc}
\usepackage{changepage}
\usepackage{hyperref,cite,url,breakurl}
\usepackage{mathtools,xparse,MnSymbol}
\usepackage{multirow}
\usepackage{stackrel}
\allowdisplaybreaks

\begin{document}
\baselineskip 0.6cm

\def\bra#1{\langle #1 |}
\def\ket#1{| #1 \rangle}
\def\inner#1#2{\langle #1 | #2 \rangle}
\def\brac#1{\llangle #1 \|}
\def\ketc#1{\| #1 \rrangle}
\def\innerc#1#2{\llangle #1 \| #2 \rrangle}
\def\Lbra#1{\bigl\langle #1 \bigr|}
\def\Lket#1{\bigl| #1 \bigr\rangle}
\def\Linner#1#2{\bigl\langle #1 \big| #2 \bigr\rangle}
\def\Lbrac#1{\bigl\llangle #1 \bigr\|}
\def\Lketc#1{\bigl\| #1 \bigr\rrangle}
\def\Linnerc#1#2{\bigl\langle #1 \big| #2 \bigr\rangle}

\def\lg{\mathrel{\lower2.5pt\vbox{\lineskip=0pt\baselineskip=0pt
           \hbox{$<$}\hbox{$>$}}}}

\def\smb{{\scalebox{0.4}{$-\beta$}}}
\def\ssmb{{\scalebox{0.3}{$-\beta$}}}
\def\spb{{\scalebox{0.4}{$+\beta$}}}
\def\sspb{{\scalebox{0.3}{$+\beta$}}}
\def\smg{{\scalebox{0.4}{$-\gamma$}}}
\def\ssmg{{\scalebox{0.3}{$-\gamma$}}}
\def\spg{{\scalebox{0.4}{$+\gamma$}}}
\def\sspg{{\scalebox{0.3}{$+\gamma$}}}
\def\spmb{{\scalebox{0.4}{$\pm\beta$}}}
\def\sspmb{{\scalebox{0.3}{$\pm\beta$}}}
\def\spmg{{\scalebox{0.4}{$\pm\gamma$}}}
\def\sspmg{{\scalebox{0.3}{$\pm\gamma$}}}
\def\smbmg{{\scalebox{0.4}{$-\beta\!-\!\gamma$}}}
\def\ssmbmg{{\scalebox{0.3}{$-\beta\!-\!\gamma$}}}
\def\spbpg{{\scalebox{0.4}{$+\beta\!+\!\gamma$}}}
\def\sspbpg{{\scalebox{0.3}{$+\beta\!+\!\gamma$}}}
\def\smbpg{{\scalebox{0.4}{$-\beta\!+\!\gamma$}}}
\def\ssmbpg{{\scalebox{0.3}{$-\beta\!+\!\gamma$}}}
\def\spbmg{{\scalebox{0.4}{$+\beta\!-\!\gamma$}}}
\def\sspbmg{{\scalebox{0.3}{$+\beta\!-\!\gamma$}}}
\newcommand{\drawsquare}[2]{\hbox{%
\rule{#2pt}{#1pt}\hskip-#2pt
\rule{#1pt}{#2pt}\hskip-#1pt
\rule[#1pt]{#1pt}{#2pt}}\rule[#1pt]{#2pt}{#2pt}\hskip-#2pt
\rule{#2pt}{#1pt}}
\newcommand{\vev}[1]{ \langle {#1} \rangle }

\makeatletter
\newcommand{\subalign}[1]{%
  \vcenter{%
    \Let@ \restore@math@cr \default@tag
    \baselineskip\fontdimen10 \scriptfont\tw@
    \advance\baselineskip\fontdimen12 \scriptfont\tw@
    \lineskip\thr@@\fontdimen8 \scriptfont\thr@@
    \lineskiplimit\lineskip
    \ialign{\hfil$\m@th\scriptstyle##$&$\m@th\scriptstyle{}##$\hfil\crcr
      #1\crcr
    }%
  }%
}
\makeatother

\begin{titlepage}

\begin{flushright}
\end{flushright}

\vskip 1.2cm

\begin{center}
{\Large \bf Black Hole Interior in Unitary Gauge Construction}

\vskip 0.7cm

{\large Yasunori Nomura}

\vskip 0.5cm

{\it Berkeley Center for Theoretical Physics, Department of Physics,\\
  University of California, Berkeley, CA 94720, USA}

\vskip 0.2cm

{\it Theoretical Physics Group, Lawrence Berkeley National Laboratory, 
 Berkeley, CA 94720, USA}

\vskip 0.2cm

{\it Kavli Institute for the Physics and Mathematics of the Universe 
 (WPI),\\
 UTIAS, The University of Tokyo, Kashiwa, Chiba 277-8583, Japan}

\vskip 0.8cm

\abstract{A quantum system with a black hole accommodates two widely different, though physically equivalent, descriptions. In one description, based on global spacetime of general relativity, the existence of the interior region is manifest, while understanding unitarity requires nonperturbative quantum gravity effects such as replica wormholes. The other description adopts a manifestly unitary, or holographic, description, in which the interior emerges effectively as a collective phenomenon of fundamental degrees of freedom.

In this paper we study the latter approach, which we refer to as the unitary gauge construction. In this picture, the formation of a black hole is signaled by the emergence of a surface (stretched horizon) possessing special dynamical properties:\ quantum chaos, fast scrambling, and low energy universality. These properties allow for constructing interior operators, as we do explicitly, without relying on details of microscopic physics. A key role is played by certain coarse modes in the zone region (hard modes), which determine the degrees of freedom relevant for the emergence of the interior.

We study how the interior operators can or cannot be extended in the space of microstates and analyze irreducible errors associated with such extension. This reveals an intrinsic ambiguity of semiclassical theory formulated with a finite number of degrees of freedom. We provide an explicit prescription of calculating interior correlators in the effective theory, which describes only a finite region of spacetime. We study the issue of state dependence of interior operators in detail and discuss a connection of the resulting picture with the quantum error correction interpretation of holography.}

\end{center}
\end{titlepage}

\tableofcontents

\section{Introduction}
\label{sec:intro}

The quantum mechanics of a black hole~\cite{Bekenstein:1973ur,Hawking:1974sw} has been a confusing subject.
On one hand, a na\"{\i}ve application of the semiclassical method leads to a violation~\cite{Hawking:1976ra} of unitarity, a fundamental principle of quantum mechanics.
On the other hand, a fully quantum mechanical treatment of the problem, such as the one based on the AdS/CFT correspondence~\cite{Maldacena:1997re}, indicates that unitarity is preserved, while it seems to be at odds with the existence of the black hole interior~\cite{Almheiri:2012rt}, a consequence of the equivalence principle of general relativity.

There are two qualitatively different, though physically equivalent, approaches to this problem, which should not be conflated with each other~\cite{Langhoff:2020jqa}.
One is to start with a global description of spacetime in general relativity.
The quantization then involves hypersurfaces that go through both the interior and exterior of the black hole, such as nice slices~\cite{Lowe:1995ac}.
In this treatment, the existence of the interior is evident by construction, and the challenge is to see how unitarity is preserved~\cite{Hawking:1976ra,Mathur:2009hf}.
This issue has recently been addressed~\cite{Penington:2019npb,Almheiri:2019psf,Almheiri:2019hni} using technologies~\cite{Ryu:2006bv,Hubeny:2007xt,Faulkner:2013ana,Engelhardt:2014gca} developed in the study of holography, where it was shown that Hawking radiation emitted from the black hole obeys the Page curve~\cite{Page:1993wv}, which is a signature of unitary evolution.
In simple (mostly lower dimensional) models, the origin of the discrepancy from Hawking's result~\cite{Hawking:1976ra} was identified:\ the nonperturbative effect called replica wormholes~\cite{Penington:2019kki,Almheiri:2019qdq}.
A pleasant surprise was that addressing the issue did not require a detailed knowledge of microscopic physics of quantum gravity.

The other approach to the problem is to begin with a manifestly unitary description.
This corresponds to viewing the black hole from a distance.
In this view, an object falling toward the black hole never crosses the horizon at the classical level, due to infinite time delay.
At the quantum level, the horizon is ``stretched'' to a timelike surface called the stretched horizon~\cite{Susskind:1993if}, on which the local (Tolman) Hawking temperature becomes the string scale.
Since the intrinsic scale of the dynamics on the stretched horizon is the string scale, it is consistent to assume, as implied by the AdS/CFT correspondence, that the physics there---and hence the black hole evolution---is unitary, despite Hawking's conclusion based on the semiclassical analysis.
In other words, one may view that the degrees of freedom outside (and on) the stretched horizon comprise the entirety of the system~\cite{tHooft:1990fkf,Susskind:1993if}.%
\footnote{We may refer to this situation either as the entire degrees of freedom being ``outside and on the stretched horizon'' or simply as ``outside the stretched horizon.''
 In the rest of the paper, we adopt the latter for brevity.}
In this picture, the challenge is to understand how a description based on near empty interior spacetime emerges~\cite{Almheiri:2012rt,Almheiri:2013hfa,Marolf:2013dba}.
In particular, we must understand why such a description applies only to the stretched horizon; after all, from the viewpoint of quantum information flow, the stretched horizon is not too different from the surface of regular material such as a piece of coal.
Our concern in this paper is this second approach, which we refer to as the ``unitary gauge construction''~\cite{Langhoff:2020jqa}.

The fact that two very different constructions lead to the same physics is a manifestation of nonperturbative gauge redundancies of a gravitational theory, which are much larger than the standard diffeomorphism~\cite{Marolf:2020xie,McNamara:2020uza} and relate even spaces with different topologies~\cite{Marolf:2020xie,Jafferis:2017tiu}.
From the quantum gravity point of view, what is special about a black hole---or more generally the system with a horizon---is that an appropriate treatment of these redundancies is vital in obtaining the correct physics.
The unitary gauge construction deals with the issue in the most intuitive way:\ the resulting picture is mostly local, although it leaves some residual nonlocality.

\subsubsection*{Unitary gauge construction}

The gist of this paper is to present the unitary gauge construction of the black hole interior as explicitly as possible and analyze its salient features.
Our construction builds on ideas and techniques introduced earlier by this and other authors~\cite{Papadodimas:2012aq,Papadodimas:2013jku,Papadodimas:2015jra,Verlinde:2012cy,Verlinde:2013qya,Nomura:2012ex,Maldacena:2013xja,Nomura:2018kia,Nomura:2019qps,Nomura:2019dlz}; however, their detailed implementations are different.
Below we discuss these earlier works, highlighting differences from ours.

A construction of interior operators based on the doubled Hilbert space structure was considered in a seminal work by Papadodimas and Raju~\cite{Papadodimas:2012aq,Papadodimas:2013jku,Papadodimas:2015jra}.
According to their prescription, the degrees of freedom that are identified as those in the first exterior of the effective two-sided black hole geometry (the region existing in the original one-sided geometry) increase as the black hole evaporates; in particular, Hawking radiation emitted earlier composes degrees of freedom in the first exterior.
On the other hand, in the construction described here, the number of degrees of freedom composing the first exterior (hard modes) decreases as the evaporation progresses~\cite{Nomura:2018kia,Nomura:2019qps,Nomura:2019dlz}; in particular, early Hawking radiation is identified as a part of the degrees of freedom in the {\it second} exterior, the region mirror to the zone of the one-sided black hole.
This leads to different solutions to some of the firewall problems, especially the $[E,\tilde{B}] \neq 0$ paradox of Ref.~\cite{Almheiri:2013hfa}.%
\footnote{A construction similar to that of Refs.~\cite{Papadodimas:2012aq,Papadodimas:2013jku,Papadodimas:2015jra} was considered in Refs.~\cite{Verlinde:2012cy,Verlinde:2013qya}, which applies only to a young black hole because of the lack of involvement of early Hawking radiation.
 The dependence of interior operators on the microstate of the system, a salient feature of the construction of Refs.~\cite{Papadodimas:2012aq,Papadodimas:2013jku,Papadodimas:2015jra}, was also noted in Ref.~\cite{Nomura:2012ex} but without an explicit operator construction.}

The idea that the construction of interior operators involves early Hawking radiation was promoted in Ref.~\cite{Maldacena:2013xja}.
A specific realization of the idea, however, is different here.
In contrast to the picture laid out in Ref.~\cite{Maldacena:2013xja}, the second exterior of the effective two-sided geometry arises primarily from degrees of freedom directly associated with the black hole (soft modes)~\cite{Nomura:2018kia,Nomura:2019qps,Nomura:2019dlz}, and the involvement of Hawking radiation is indirect (although it is significant for an old black hole, i.e.\ a black hole that is nearly maximally entangled with the rest of the system).
In particular, the structure of entanglement is not bipartite between the first exterior and early Hawking radiation degrees of freedom as envisioned in Ref.~\cite{Maldacena:2013xja}.

\begin{figure}[t]
\begin{center}
  \includegraphics[height=6.3cm]{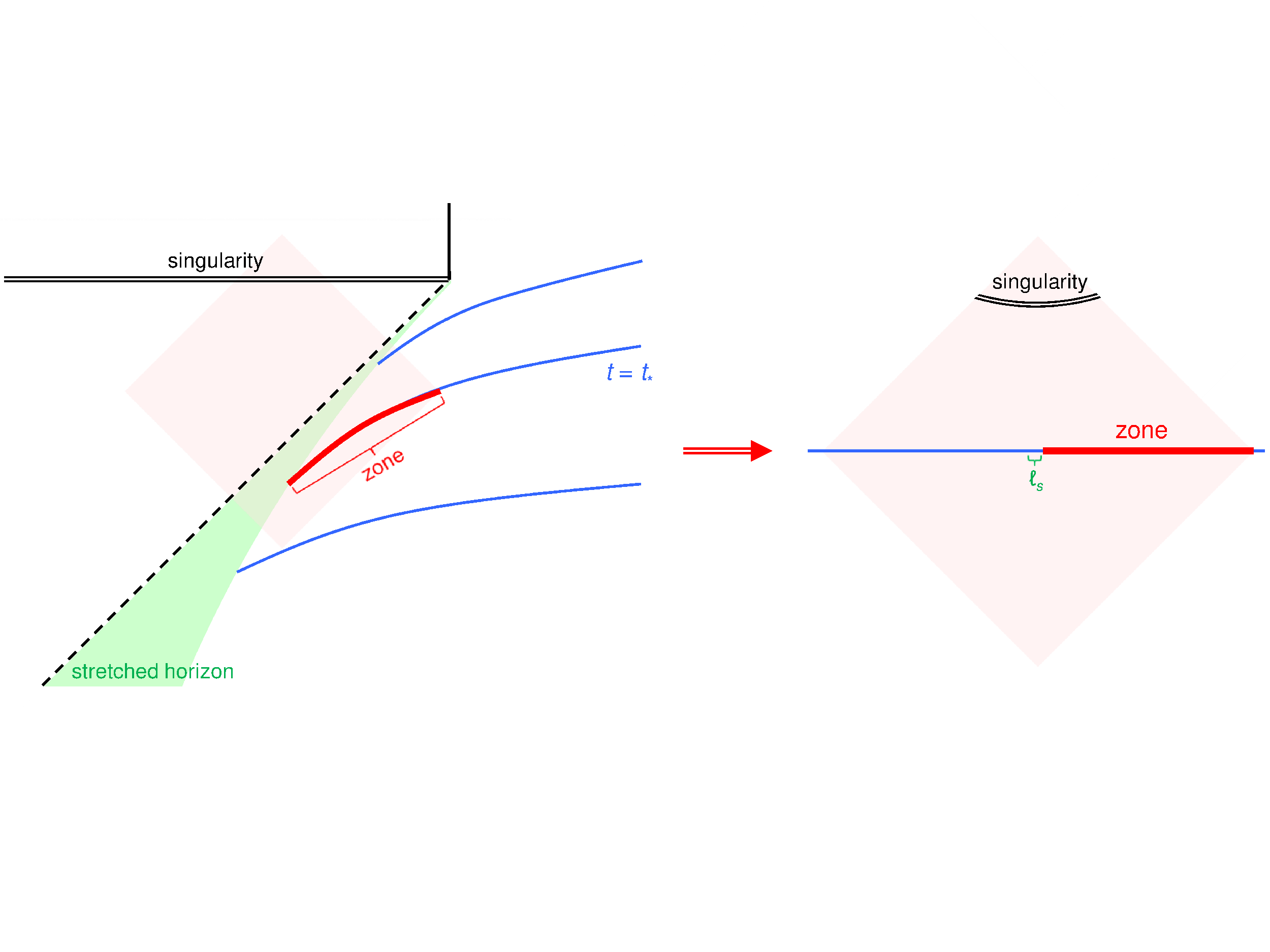}
\end{center}
\caption{The effective theory of the interior can be erected at a boundary time $t_*$ in the original unitary theory which has a one-sided black hole.
 It has an effective two-sided black hole geometry and describes physics in the causal domain of the union of the zone and its mirror region on the spatial hypersurface corresponding to the $t_*$ hypersurface.
 The effective theory is intrinsically semiclassical and cannot describe dynamics occurring above the string scale $1/l_{\rm s}$ (although space below $l_{\rm s}$ exists to accommodate an object kinematically squeezed by a large Lorentz boost).}
\label{fig:two-sided}
\end{figure}
A description of the black hole interior obtained by our construction applies only to a limited spacetime region---the causal domain of the union of the zone and its mirror on the spatial hypersurface determined by the time at which the description is erected~\cite{Nomura:2018kia,Nomura:2019qps} (see Fig.~\ref{fig:two-sided}).
In order to cover a larger portion of the black hole interior, we must use multiple descriptions erected at different times.
This provides a specific realization of the idea of black hole complementarity~\cite{Susskind:1993if,Susskind:1993mu}.
Unlike what is envisioned in some of the early works (e.g.\ Ref.~\cite{Kiem:1995iy}), however, the exterior and interior descriptions are not related by a unitary transformation; rather, the latter emerges through coarse graining and is intrinsically semiclassical.
In fact, this coarse graining is the origin of the apparent uniqueness of the infalling vacuum, despite the existence of exponentially many black hole microstates.

Our construction essentially follows that of Refs.~\cite{Nomura:2018kia,Nomura:2019qps,Nomura:2019dlz}.
In particular, microscopic operators describing the interior are the same as those in Ref.~\cite{Nomura:2019dlz}.
However, our analyses on an intrinsic ambiguity of semiclassical theory and state (in)dependence of interior operators are different, which supersede the discussion of Refs.~\cite{Nomura:2019dlz,Langhoff:2020jqa} about the issues.

As discussed in Refs.~\cite{Langhoff:2020jqa,Nomura:2018kia,Nomura:2019qps,Nomura:2019dlz}, what distinguishes the stretched horizon from other, regular material surfaces is its kinematic feature of an exponentially large density of states as well as its special dynamical properties---maximal quantum chaos~\cite{Maldacena:2015waa}, fast scrambling~\cite{Hayden:2007cs,Sekino:2008he}, and the lack of a feature discriminating low energy species, such as global symmetries~\cite{Banks:2010zn,Harlow:2018jwu}.
The fact that the properties of the ultraviolet (UV) dynamics in one description are related to the existence of large/infrared (IR) interior spacetime in another description is an example of UV-IR relations pervasive in quantum gravity.
Note that the criterion necessary for a surface to be a stretched horizon is stronger than that for regular thermalization occurring around us; in particular, it must exhibit ``universal thermalization'' applicable throughout all the low energy species.
Such a strong universality presumably arises only as a result of the string dynamics.

We stress that the prescription of obtaining interior operators described here does not require a detailed knowledge of microscopic dynamics of quantum gravity.
This insensitivity to the microscopic physics puts the unitary gauge construction on an equal footing with the approach~\cite{Penington:2019npb,Almheiri:2019psf,Almheiri:2019hni,Penington:2019kki,Almheiri:2019qdq} based on the global spacetime picture.

\subsubsection*{Outline of the paper}

In Section~\ref{sec:unitary}, we discuss how a black hole and its evolution is described in the unitary gauge construction when the black hole is viewed from a distance.
This corresponds to the boundary description in holography, and the evolution of the system is unitary.

In Section~\ref{sec:operators}, we discuss in detail how operators used in describing the black hole interior are constructed out of modes in the distant description.
A key role is played by what is called the hard modes:\ a type of coarse modes in the black hole zone which semiclassical theory describes.
The degrees of freedom necessary to construct the interior are chosen by using them as an anchor.
They consist of both black hole micro degrees of freedom, called soft modes, and the degrees of freedom entangled with the soft and hard modes.
We discuss how the infalling operators constructed in this way can be promoted to act more globally in the space of vacuum microstates and derive the level of the precision preserved by such a promotion.

In Section~\ref{sec:eff-theory}, we discuss the effective theory of the interior in the unitary gauge construction.
This theory can be erected at each boundary time using suitably promoted infalling operators.
It is intrinsically semiclassical, describing the dynamics of hard-mode excitations and their fate after crossing the horizon.
We analyze the irreducible error resulting from the process of erecting the theory, which is viewed as an intrinsic ambiguity of semiclassical theory formulated with a finite number of degrees of freedom.
We also present an explicit prescription for computing interior correlators in the effective theory.
We then discuss state dependence of infalling operators and a connection of the resulting picture with the quantum error correction interpretation of holography.
We finally comment on a young black hole, for which infalling operators can be constructed only out of the soft modes, and the Minkowski limit.

Section~\ref{sec:concl} is devoted to the conclusion.

Throughout the paper, we focus on a spherically symmetric, non-near extremal black hole in asymptotically flat or AdS spacetime, although we expect that a similar construction applies to other black holes as well.
We take the Schr\"{o}dinger picture of quantum mechanics unless otherwise stated and adopt natural units $c = \hbar = 1$.

\section{Black Hole Evolution as Viewed from the Exterior}
\label{sec:unitary}

The unitary gauge construction adopts a view that when a black hole is described in a ``distant reference frame,'' the degrees of freedom outside the horizon comprise the entire system.
Namely, these degrees of freedom evolve unitarily~\cite{tHooft:1990fkf,Susskind:1993if,Page:1993wv} under time evolution associated with an external observer, or boundary time evolution in holography.%
\footnote{In this paper, we assume that the external observer is located sufficiently far from the black hole and refer to the time associated with them as boundary time, even if they may not be in the true asymptotic region.}
At the classical level, an object falling toward the black hole never reaches the horizon because of infinite time delay caused by a diverging gravitational redshift.
At the quantum level, the horizon becomes a timelike surface called the stretched horizon~\cite{Susskind:1993if}, on which the local Hawking temperature becomes the string scale and where a falling object reaches in a finite boundary time.
The basic tenet of the unitary gauge construction is that the full quantum degrees of freedom consist of those outside the stretched horizon.

In general, the state of the system at a given boundary time $t$ can be expanded in terms of the states of the black hole in question.%
\footnote{The boundary time can be related to a bulk equal-time hypersurface through a gauge fixing procedure, for example by the ``holographic slice'' prescription of Refs.~\cite{Nomura:2018kji,Murdia:2020iac}.}
A state of the black hole can be labeled by two sets of indices:\ indices specifying the macroscopic properties of the black hole and additional indices necessary to specify the black hole microstate uniquely.
(The precise meaning of the black hole microstate is discussed later.)
For a non-rotating and uncharged black hole, the state of the system can be written as
\begin{equation}
  \ket{\Psi(t)} = \sum_M \sum_{A_M} \sum_I d_{M A_M I}(t) \ket{\Psi_{A_M, I}(M)}.
\label{eq:Psi-t_gen}
\end{equation}
Here, $M$ is the mass of the black hole specified up to the precision that can be discriminated by a temporal observer (determined by the uncertainty principle):
\begin{equation}
  \Delta = O(2\pi T_{\rm H}),
\end{equation}
where $T_{\rm H}$ is the Hawking temperature, and $A_M$ is the additional index needed to specify the black hole microstate uniquely for each $M$.
The index $I$ labels all the excitations over the semiclassical black hole vacuum (i.e.\ a state that has only the black hole in question) in regions near and far from the black hole.%
\footnote{For example, if there is another black hole beyond the one we focus on, we treat it as an excitation over the semiclassical vacuum and label it by $I$.
 We do this simply to focus on the black hole in question.}

Below, we concentrate on a branch with a fixed $M$ at each time $t$ and, accordingly, drop the subscript $M$ from the index $A$:
\begin{equation}
  \ket{\Psi_{A_M,I}(M)} \rightarrow \ket{\Psi_{A,I}(M)}.
\label{eq:A_M-A}
\end{equation}
Including other branches into consideration is straightforward.

\subsubsection*{Vacuum microstates}

In general, there are three classes of degrees of freedom associated with a black hole:\ hard modes, soft modes, and far modes (radiation)~\cite{Nomura:2018kia,Nomura:2019qps,Nomura:2019dlz}.
Hard and soft modes are the degrees of freedom near the black hole, often called the zone or black hole atmosphere region:
\begin{equation}
  r_{\rm s} \leq r \leq r_{\rm z},
\end{equation}
where $r$ is the area radius with $r_{\rm s}$ and $r_{\rm z}$ representing the location of the stretched horizon and the edge of the zone, respectively.
Hard modes are the coarse degrees of freedom in this region, which have frequencies $\omega$ and gaps among them $\varDelta\omega$ (sufficiently) larger than $\Delta$,%
\footnote{When we refer to energy, frequency, and so on, we mean those as measured in the far (or asymptotic) region, unless otherwise stated.}
while the soft modes have smaller frequencies; far modes are the degrees of freedom outside the zone, $r > r_{\rm z}$.
The dynamics of the hard and far modes are described by semiclassical theory, while the soft modes (and a part of the far modes) comprise black hole microstates.
More detailed discussion about these modes with various examples can be found in Ref.~\cite{Langhoff:2020jqa}.

Let us consider the situation in which there are no excitations beyond those directly associated with the existence of the black hole, which we refer to as the system being in the (semiclassical) black hole vacuum and represent by $I = 0$.
This does not mean that hard modes, soft modes, and far modes are all in their ground states.
Because of entanglement between these modes and the energy constraint coming from the fact that the black hole has mass $M$, a black hole vacuum microstate is given by~\cite{Nomura:2018kia,Nomura:2019qps,Nomura:2019dlz}
\begin{equation}
  \ket{\Psi_{A,0}(M)} = \sum_n \sum_{i_n = 1}^{e^{S_{\rm bh}(M-E_n)}} 
    \sum_{a = 1}^{e^{S_{\rm rad}}} c^A_{n i_n a} \ket{\{ n_\alpha \}} 
    \ket{\psi^{(n)}_{i_n}} \ket{\phi_a}.
\label{eq:sys-state}
\end{equation}
Here, $\ket{\{ n_\alpha \}}$, $\ket{\psi^{(n)}_{i_n}}$, and $\ket{\phi_a}$ are orthonormal states of the hard modes, soft modes, and far modes, respectively:
\begin{equation}
  \inner{\{ m_\alpha \}}{\{ n_\alpha \}} = \delta_{m n},
\qquad
  \inner{\psi^{(m)}_{i_m}}{\psi^{(n)}_{j_n}} = \delta_{m n} \delta_{i_m j_n},
\qquad
  \inner{\phi_a}{\phi_b} = \delta_{ab},
\label{eq:orthonorm}
\end{equation}
where $n \equiv \{ n_\alpha \}$ represents the set of all occupation numbers $n_\alpha$ ($\geq 0$) for the hard modes, which are labeled by $\alpha$ (collectively denoting the species, frequency, and angular-momentum quantum numbers),%
\footnote{Because of the energy uncertainty of order $\Delta$, the inner products of hard and soft modes in Eq.~(\ref{eq:orthonorm}) may have exponentially suppressed corrections of order $\inner{\{ m_\alpha \}}{\{ n_\alpha \}} \sim \inner{\psi^{(m)}_{i_m}}{\psi^{(n)}_{j_n}} \sim e^{-|E_m-E_n|/\Delta}$ (although it may be possible to avoid this by defining hard modes using a smoothing function in frequency space which damps rapidly outside the window of order $\Delta$).
 These corrections, if any, are typically very small for a semiclassical object, $E_{m,n} \gg \Delta$, so we ignore them in the rest of the paper.}
$E_n$ is the energy of the hard mode state $\ket{\{ n_\alpha \}}$, and $S_{\rm bh}(E)$ is the Bekenstein-Hawking entropy density at energy $E$; see Ref.~\cite{Langhoff:2020jqa} for a more detailed description of these states.

For the analysis of this paper, we envision the simplest setup in which the relevant components of the far modes consist of Hawking radiation emitted earlier from the black hole.
In general, they must involve all the degrees of freedom entangled with the hard and soft modes.
Such entanglement can be generated through direct or indirect interactions of these degrees of freedom with the black hole or matter forming it.
Including this effect does not affect the analysis in this paper; it simply requires a reinterpretation of $\ket{\phi_a}$.

The index $A$ of $\ket{\Psi_{A,0}(M)}$ labels microstates specified by the coefficients $c^A_{n i_n a}$.
The number of independent microstates $e^{S_{\rm tot}}$ is determined by the coarse-grained entropies of the soft modes $S_{\rm bh}(E_{\rm soft})$ and the far modes/early radiation $S_{\rm rad}$, and we let the index $A$ label the orthonormal basis states (of an arbitrary basis):%
\footnote{Recall that $\ket{\Psi_{A,0}(M)}$ represent microstates of the soft mode and radiation with the black hole put in the semiclassical vacuum, so that a generic state in the Hilbert space of dimension $e^{S_{\rm tot}}$ has the black hole of mass $M$.
Note that since black hole evaporation is a thermodynamically irreversible process~\cite{Zurek:1982zz,Page:1983ug}, most of these microstates do not become a state with a larger black hole in empty space when evolved backward in time---there is some junk radiation around it.
This, however, does not change the fact that there are $e^{S_{\rm tot}}$ independent microstates relevant for the discussion here.}
\begin{equation}
  A = 1,\cdots,e^{S_{\rm tot}},
\end{equation}
where
\begin{equation}
  e^{S_{\rm tot}} \equiv \sum_n e^{S_{\rm bh}(M-E_n)} e^{S_{\rm rad}} = z\, e^{S_{\rm bh}(M)+S_{\rm rad}}
\end{equation}
with
\begin{equation}
  z \equiv \sum_n e^{-\frac{E_n}{T_{\rm H}}}.
\label{eq:def-z}
\end{equation}
With this convention, the coefficients $c^A_{n i_n a}$ satisfy
\begin{equation}
  \sum_n \sum_{i_n = 1}^{e^{S_{\rm bh}(M-E_n)}} \sum_{a = 1}^{e^{S_{\rm rad}}} 
    c^{A*}_{n i_n a} c^{B}_{n i_n a} = \delta_{AB}.
\label{eq:norm}
\end{equation}

The spatial distribution of the soft modes, which carry the energy and entropy of the black hole, is determined by the local Hawking temperature
\begin{equation}
  T_{\rm loc}(r) = \frac{T_{\rm H}}{\sqrt{-g_{tt}(r)}}.
\end{equation}
This distribution is strongly peaked toward the stretched horizon, where the local temperature reaches the string scale.
Since the invariant dynamical scale among the soft modes is of the order of the local temperature, their internal dynamics is controlled by the microscopic dynamics of quantum gravity and cannot be described by a low energy theory; indeed, we expect that it is nonlocal in the spatial directions along the horizon~\cite{Hayden:2007cs,Sekino:2008he}.
Nevertheless, it is widely believed that this dynamics exhibits certain characteristic behaviors; in particular, it is maximally quantum chaotic~\cite{Maldacena:2015waa}, fast scrambling~\cite{Hayden:2007cs,Sekino:2008he}, and does not have a feature discriminating low energy species beyond their spacetime and gauge properties~\cite{Banks:2010zn,Harlow:2018jwu}.
As argued in Refs.~\cite{Nomura:2019qps,Nomura:2019dlz}, these are critical ingredients that distinguish the stretched horizon from normal material surfaces, leading to near empty interior spacetime.

Specifically, the dynamical properties described above imply that the coefficients $c^A_{n i_n a}$ in Eq.~(\ref{eq:sys-state}) have the statistical properties
\begin{equation}
  \vev{c^A_{n i_n a}} = 0,
\qquad
  \sqrt{\vev{|c^A_{n i_n a}|^2}} = \frac{1}{e^{\frac{1}{2}S_{\rm tot}}},
\label{eq:c-size}
\end{equation}
where $\vev{\cdots}$ represents an ensemble average over $(i_n,a)$, and that the phases of $c^A_{n i_n a}$'s are distributed uniformly.
In fact, this configuration is reached quickly, within the scrambling time $t_{\rm scr}$ of order $(1/2\pi T_{\rm H})\ln S_{\rm bh}(M)$.
With Eq.~(\ref{eq:c-size}), we can trace out the soft modes, obtaining the thermal density matrix for the hard (i.e.\ semiclassical) modes
\begin{equation}
  {\rm Tr}_{\rm soft} \ket{\Psi_{A,0}(M)} \bra{\Psi_{A,0}(M)}
  = \frac{1}{z} \sum_n e^{-\frac{E_n}{T_{\rm H}}} \ket{\{ n_\alpha \}} \bra{\{ n_\alpha \}} \otimes \rho_\phi,
\label{eq:rho_HR}
\end{equation}
where $\rho_\phi$ is an $n$-independent reduced density matrix for the far modes; fractional corrections to the coefficients of $\ket{\{ n_\alpha \}} \bra{\{ n_\alpha \}}$ and the matrix elements of $\rho_\phi$ (which are in general $n$~dependent) are only of order $e^{-\frac{1}{2}S_{\rm bh}(M-E_n)}$.
This is indeed the origin of the thermality of the black hole atmosphere in semiclassical theory~\cite{Nomura:2018kia}.

\subsubsection*{Excitations in the zone}

Suppose that the state of the system is given by Eq.~(\ref{eq:sys-state}) at a boundary time $t$.
Field operators in the zone in the semiclassical theory are then expanded in terms of annihilation and creation operators for the hard modes
\begin{align}
  b_\gamma &= \sum_n \sqrt{n_\gamma}\, 
    \ket{\{ n_\alpha - \delta_{\alpha\gamma} \}} \bra{\{ n_\alpha \}},
\label{eq:ann}\\*
  b_\gamma^\dagger &= \sum_n \sqrt{n_\gamma + 1}\, 
    \ket{\{ n_\alpha + \delta_{\alpha\gamma} \}} \bra{\{ n_\alpha \}}.
\label{eq:cre}
\end{align}
Since the semiclassical theory is not sensitive to the microstate of the black hole and is local in spacetime, these operators do not act on soft or far mode degrees of freedom.

Acting these operators on vacuum microstate $\ket{\Psi_{A,0}(M)}$, we find
\begin{align}
  b_\gamma\, \ket{\Psi_{A,0}(M)} &= \sum_n \sqrt{n_\gamma} 
    \sum_{i_n = 1}^{e^{S_{\rm bh}(M-E_n)}} \sum_{a = 1}^{e^{S_{\rm rad}}} c^A_{n i_n a} 
    \ket{\{ n_\alpha - \delta_{\alpha\gamma} \}} \ket{\psi^{(n)}_{i_n}} \ket{\phi_a},
\label{eq:b-exc}\\*
  b_\gamma^\dagger\, \ket{\Psi_{A,0}(M)} &= \sum_n \sqrt{n_\gamma + 1} 
    \sum_{i_n = 1}^{e^{S_{\rm bh}(M-E_n)}} \sum_{a = 1}^{e^{S_{\rm rad}}} c^A_{n i_n a} 
    \ket{\{ n_\alpha + \delta_{\alpha\gamma} \}} \ket{\psi^{(n)}_{i_n}} \ket{\phi_a}.
\label{eq:bd-exc}
\end{align}
Note that a state obtained by acting these operators on $\ket{\Psi_{A,0}(M)}$ cannot be viewed as a vacuum state.
First, it generically breaks the symmetry of the black hole spacetime; for example, it may have an excitation localized in the angular directions.
Second, excitations of such a state are generically nonuniversal; namely, only specific low energy species are excited.
Finally, even if these features were disregarded, states obtained by acting $b_\gamma$'s and/or $b_\gamma^\dagger$'s on $\ket{\Psi_{A,0}(M)}$ do not possess properties needed to play the role of a vacuum state in the construction of Refs.~\cite{Nomura:2018kia,Nomura:2019qps,Nomura:2019dlz}; in particular, they do not lead to the reduced density matrix of the form in Eq.~(\ref{eq:rho_HR}).%
\footnote{The Born rule problem of Ref.~\cite{Marolf:2015dia} does not apply to our construction, since the excited states obtained in this way are atypical in the microscopic Hilbert space~\cite{Nomura:2019dlz,Langhoff:2020jqa}.
 The frozen vacuum problem of Ref.~\cite{Bousso:2013ifa} does not apply either; see Section~\ref{sec:eff-theory} for more details.}

From Eqs.~(\ref{eq:b-exc},~\ref{eq:bd-exc}), we find
\begin{align}
  \bra{\Psi_{A,0}(M)} b_\beta^\dagger b_\gamma \ket{\Psi_{B,0}(M)} 
  &= \delta_{\beta\gamma} \sum_n \sum_{i_n = 1}^{e^{S_{\rm bh}(M-E_n)}} 
    \sum_{a = 1}^{e^{S_{\rm rad}}} n_\gamma\, c^{A*}_{n i_n a} c^{B}_{n i_n a},
\label{eq:non-ortho-1}\\*
  \bra{\Psi_{A,0}(M)} b_\beta b_\gamma^\dagger \ket{\Psi_{B,0}(M)} 
  &= \delta_{\beta\gamma} \sum_n \sum_{i_n = 1}^{e^{S_{\rm bh}(M-E_n)}} 
    \sum_{a = 1}^{e^{S_{\rm rad}}} (n_\gamma + 1)\, c^{A*}_{n i_n a} c^{B}_{n i_n a}.
\label{eq:non-ortho-2}
\end{align}
These are not proportional to $\delta_{AB}$ in general, although deviations from it are suppressed exponentially by a factor of $e^{-\frac{1}{2}\{ S_{\rm bh}(M) + S_{\rm rad}\} }$.
In particular, this implies that the commutator $[b_\beta, b_\gamma^\dagger]$ is not $\delta_{\beta\gamma}$ as an operator at the microscopic level, although its vacuum expectation values satisfy
\begin{equation}
  \bra{\Psi_{A,0}(M)} [b_\beta, b_\gamma^\dagger] \ket{\Psi_{B,0}(M)} 
  = \delta_{\beta\gamma} \sum_n \sum_{i_n = 1}^{e^{S_{\rm bh}(M-E_n)}} 
    \sum_{a = 1}^{e^{S_{\rm rad}}} c^{A*}_{n i_n a} c^{B}_{n i_n a} 
  = \delta_{\beta\gamma} \delta_{AB}.
\end{equation}

States described here evolve unitarily under boundary time evolution.
The time evolution of (a superposition of) microstates of the form of Eq.~(\ref{eq:sys-state})---particularly under the Hawking emission process---was discussed in Refs.~\cite{Nomura:2018kia,Nomura:2014woa}.
While a complete description of the evolution requires a microscopic theory of quantum gravity, we can write down an evolution equation leaving the coefficients $c^A_{n i_n a}$ unspecified.
In particular, Hawking emission occurs through soft modes at the edge of the zone.
At the level of ignoring Hawking emission and its backreaction, the dynamics of semiclassical objects in the near horizon region is described by the Hamiltonian
\begin{equation}
  H = \sum_\gamma \omega_\gamma b_\gamma^\dagger b_\gamma + H_{\rm int}\bigl( \{ b_\gamma \}, \{ b_\gamma^\dagger \} \bigr),
\label{eq:H}
\end{equation}
where $\omega_\gamma$ is the frequency of mode $\gamma$.
In particular, the evolution of a state by a boundary time $\varDelta t$ is given by the time evolution operator $e^{-i H \varDelta t}$.

\section{Quantum Operators for the Interior}
\label{sec:operators}

Since all the degrees of freedom in the unitary gauge construction exist outside the stretched horizon, we have to find the degrees of freedom that effectively describe the interior within these exterior degrees of freedom.
This boils down to identifying the degrees of freedom that can play the role of the ``second exterior'' of an analytically extended two-sided black hole~\cite{Papadodimas:2012aq}.
As discussed in Refs.~\cite{Nomura:2018kia,Nomura:2019qps,Nomura:2019dlz}, this can be done at each boundary time, and the degrees of freedom can be identified in the combined system of the soft and far modes.
(For a young black hole, i.e.\ a black hole in which the soft modes are not maximally entangled with the far modes, the soft modes alone can provide the necessary degrees of freedom~\cite{Nomura:2019dlz}.
We will discuss this in Section~\ref{subsec:comments}.)
In this section, we elaborate on the construction of Refs.~\cite{Nomura:2018kia,Nomura:2019qps,Nomura:2019dlz} and derive formulas that are used in our later discussion of the black hole interior.

\subsection{Mirror microstates}
\label{subsec:mirror-st}

Suppose that the state of the system at a boundary time $t$ is given by Eq.~(\ref{eq:sys-state}) (possibly) with excitations of hard modes and/or far modes over it.
Following the lines of Refs.~\cite{Papadodimas:2012aq,Papadodimas:2013jku,Papadodimas:2015jra,Verlinde:2012cy,Verlinde:2013qya,Nomura:2012ex}, we define normalized {\it mirror microstates} $\ketc{\{ n_\alpha \}_A}$ as the state of the soft and far modes entangled with the hard mode state $\ket{\{ n_\alpha \}}$ in the corresponding vacuum microstate:
\begin{equation}
  \ketc{\{ n_\alpha \}_A} = \alpha^A_n \sum_{i_n = 1}^{e^{S_{\rm bh}(M-E_n)}} \sum_{a = 1}^{e^{S_{\rm rad}}} c^A_{n i_n a} \ket{\psi^{(n)}_{i_n}} \ket{\phi_a}.
\label{eq:ketc}
\end{equation}
Here, the normalization constant $\alpha^A_n$ is given by
\begin{align}
  \alpha^A_n &= \frac{1}{\sqrt{\sum_{i_n = 1}^{e^{S_{\rm bh}(M-E_n)}} \sum_{a = 1}^{e^{S_{\rm rad}}} c^{A*}_{n i_n a} c^A_{n i_n a}}}
\nonumber\\*
  &= \sqrt{z}\,\, e^{\frac{E_n}{2T_{\rm H}}} \left( 1 - \frac{1}{2}\varepsilon_n^{AA} \right),
\label{eq:alpha_nA}
\end{align}
where $z$ is given by Eq.~(\ref{eq:def-z}), and we have used statistical properties of $c^A_{n i_n a}$ to obtain the last expression, with $\varepsilon_n^{AA}$ defined below.
Note that these states are defined at the boundary time $t$, where the state of the system takes the assumed form.

The quantity $\varepsilon_n^{AA}$ in Eq.~(\ref{eq:alpha_nA}) is exponentially suppressed.
Let us define related, more general quantities by
\begin{equation}
  \varepsilon_n^{AB} \equiv z\, e^{\frac{E_n}{T_{\rm H}}} \sum_{i_n = 1}^{e^{S_{\rm bh}(M-E_n)}} \sum_{a = 1}^{e^{S_{\rm rad}}} c^{A*}_{n i_n a} c^B_{n i_n a} - \delta_{AB},
\end{equation}
of which $\varepsilon_n^{AA}$ is the special case.
These quantities satisfy the statistical properties
\begin{equation}
  \vev{\varepsilon_n^{AB}} = 0,
\qquad
  \sqrt{\vev{|\varepsilon_n^{AB}|^2}} = O\left( \frac{1}{e^{\frac{1}{2}\{ S_{\rm bh}(M-E_n)+S_{\rm rad} \}}} \right),
\label{eq:epsilon-AB}
\end{equation}
where $\vev{\cdots}$ represents an ensemble average over $A$ for $A = B$ and over $(A,B)$ for $A \neq B$.
They also obey
\begin{equation}
  (\varepsilon_n^{AB})^* = \varepsilon_n^{BA},
\qquad
  \sum_n \frac{e^{-\frac{E_n}{T_{\rm H}}}}{z} \varepsilon_n^{AB} = 0,
\label{eq:epsilon-sum}
\end{equation}
where the second equation follows from Eq.~(\ref{eq:norm}).

The mirror microstates defined above have inner products
\begin{align}
  \innerc{\{ m_\alpha \}_A}{\{ n_\alpha \}_B} 
  &= \delta_{m n}\, \alpha^A_n \alpha^B_n \sum_{i_n = 1}^{e^{S_{\rm bh}(M-E_n)}} \sum_{a = 1}^{e^{S_{\rm rad}}} c^{A*}_{n i_n a} c^{B}_{n i_n a} 
\nonumber\\*
  &= \delta_{m n}\, \eta_n^{AB},
\label{eq:inner-c}
\end{align}
where
\begin{equation}
  \eta_n^{AB} \equiv \begin{cases} 
    1 & \mbox{for } A = B \\
    \varepsilon_n^{AB} & \mbox{for } A \neq B.
  \end{cases}
\label{eq:eta}
\end{equation}
Note that $\varepsilon_n^{AB}$ are exponentially small; see Eq.~(\ref{eq:epsilon-AB}).

\subsection{Canonical mirror operators for a microstate}
\label{subsec:mirror-op}

The {\it canonical mirror operators} for microstate $A$ are defined as the ``annihilation and creation operators'' for the corresponding mirror microstates~\cite{Nomura:2018kia,Nomura:2019qps,Nomura:2019dlz}:
\begin{align}
  \tilde{b}_\gamma^A &= \sum_n \sqrt{n_\gamma}\, \Lketc{\{ n_\alpha - \delta_{\alpha\gamma} \}_A} \Lbrac{\{ n_\alpha \}_A},
\label{eq:ann-m-orig}\\*
  \tilde{b}_\gamma^{A\dagger} &= \sum_n \sqrt{n_\gamma + 1}\, \Lketc{\{ n_\alpha + \delta_{\alpha\gamma} \}_A} \Lbrac{\{ n_\alpha \}_A}.
\label{eq:cre-m-orig}
\end{align}
Given that the vacuum microstate in Eq.~(\ref{eq:sys-state}) can be written in the thermofield double form as
\begin{equation}
  \ket{\Psi_{A,0}(M)} = \frac{1}{\sqrt{z}} \sum_n e^{-\frac{E_n}{2T_{\rm H}}} \ket{\{ n_\alpha \}} \ketc{\{ n_\alpha \}_A}
\label{eq:TFD}
\end{equation}
(up to exponentially small corrections of order $\varepsilon_n^{AA}$), these operators can be viewed as the annihilation and creation operators in the zone of the second exterior region.
Note that at this point, the operators are defined only on the equal-time hypersurface of the effective two-sided black hole obtained by analytically continuing the zone of the original black hole at the boundary time $t$.
Time evolution of these operators (in the Heisenberg picture) or states of the effective two-sided theory (in the Schr\"{o}dinger picture) will be discussed in Section~\ref{sec:eff-theory}.

Products of the canonical mirror operators in Eqs.~(\ref{eq:ann-m-orig},~\ref{eq:cre-m-orig}) are given by
\begin{align}
  \tilde{b}_\beta^A \tilde{b}_\gamma^B &= \sum_n \sqrt{(n_\beta - \delta_{\beta\gamma}) n_\gamma}\,\, \eta_{n_\smg}^{AB}\, \Lketc{\{ n_\alpha - \delta_{\alpha\beta} - \delta_{\alpha\gamma} \}_A} \Lbrac{\{ n_\alpha \}_B},
\\*
  \tilde{b}_\beta^{A\dagger} \tilde{b}_\gamma^{B\dagger} &= \sum_n \sqrt{(n_\beta + 1 + \delta_{\beta\gamma}) (n_\gamma + 1)}\,\, \eta_{n_\spg}^{AB}\, \Lketc{\{ n_\alpha + \delta_{\alpha\beta} + \delta_{\alpha\gamma} \}_A} \Lbrac{\{ n_\alpha \}_B},
\\*
  \tilde{b}_\beta^A \tilde{b}_\gamma^{B\dagger} &= \sum_n \sqrt{(n_\beta + \delta_{\beta\gamma}) (n_\gamma + 1)}\,\, \eta_{n_\spg}^{AB}\, \Lketc{\{ n_\alpha - \delta_{\alpha\beta} + \delta_{\alpha\gamma} \}_A} \Lbrac{\{ n_\alpha \}_B},
\\*
  \tilde{b}_\beta^{A\dagger} \tilde{b}_\gamma^B &= \sum_n \sqrt{(n_\beta + 1 - \delta_{\beta\gamma}) n_\gamma}\,\, \eta_{n_\smg}^{AB}\, \Lketc{\{ n_\alpha + \delta_{\alpha\beta} - \delta_{\alpha\gamma} \}_A} \Lbrac{\{ n_\alpha \}_B},
\end{align}
where $n_{\pm\gamma} \equiv \{ n_\alpha \pm \delta_{\alpha\gamma} \}$.
Thus, their commutators are given by
\begin{equation}
  [\tilde{b}_\beta^A, \tilde{b}_\gamma^B] = \sum_n\, \Bigl\{ \sqrt{(n_\beta - \delta_{\beta\gamma}) n_\gamma}\,\, \eta_{n_\smg}^{AB} - \sqrt{(n_\gamma - \delta_{\beta\gamma}) n_\beta}\,\, \eta_{n_\smb}^{AB} \Bigr\}
    \Lketc{\{ n_\alpha - \delta_{\alpha\beta} - \delta_{\alpha\gamma} \}_A} \Lbrac{\{ n_\alpha \}_B},
\label{eq:bb}
\end{equation}
\begin{align}
  [\tilde{b}_\beta^{A\dagger}, \tilde{b}_\gamma^{B\dagger}] &= \sum_n\, \Bigl[ \sqrt{(n_\beta + 1 + \delta_{\beta\gamma}) (n_\gamma + 1)}\,\, \eta_{n_\spg}^{AB} - \sqrt{(n_\gamma + 1 + \delta_{\beta\gamma}) (n_\beta + 1)}\,\, \eta_{n_\spb}^{AB} \Bigr] 
\nonumber\\*
  &\qquad\qquad\times \Lketc{\{ n_\alpha + \delta_{\alpha\beta} + \delta_{\alpha\gamma} \}_A} \Lbrac{\{ n_\alpha \}_B},
\label{eq:bdbd}
\end{align}
\begin{align}
  & [\tilde{b}_\beta^A, \tilde{b}_\gamma^{B\dagger}] = \sum_n\, \Bigl[ \sqrt{(n_\beta + \delta_{\beta\gamma}) (n_\gamma + 1)}\,\, \eta_{n_\spg}^{AB} - \sqrt{(n_\gamma + 1 - \delta_{\beta\gamma}) n_\beta}\,\, \eta_{n_\smb}^{AB} \Bigr] \Lketc{\{ n_\alpha - \delta_{\alpha\beta} + \delta_{\alpha\gamma} \}_A} \Lbrac{\{ n_\alpha \}_B}
\nonumber\\*
  &\qquad\quad\,\,\,\, = \left\{ \begin{array}{l} \sum_n \Bigl[ \eta_{n_\spg}^{AB} + n_\gamma (\varepsilon_{n_\spg}^{AB} - \varepsilon_{n_\smg}^{AB}) (1-\delta_{AB}) \Bigr]\, \Lketc{\{ n_\alpha \}_A} \Lbrac{\{ n_\alpha \}_B}
  \qquad \mbox{for } \beta = \gamma
\\
  \sum_n \sqrt{n_\beta (n_\gamma + 1)}\, (\varepsilon_{n_\spg}^{AB} - \varepsilon_{n_\smg}^{AB})(1-\delta_{AB})\, \Lketc{\{ n_\alpha - \delta_{\alpha\beta} + \delta_{\alpha\gamma} \}_A} \Lbrac{\{ n_\alpha \}_B}
  \qquad \mbox{for } \beta \neq \gamma,
\end{array} \right.
\label{eq:bbd-2}
\end{align}
where $\varepsilon_{n_\spmb}^{AB}$ are exponentially small.
Interestingly, for $A = B$, we obtain
\begin{equation}
  [\tilde{b}_\beta^A, \tilde{b}_\gamma^A] = [\tilde{b}_\beta^{A\dagger}, \tilde{b}_\gamma^{A\dagger}] = 0,
\quad
  [\tilde{b}_\beta^A, \tilde{b}_\gamma^{A\dagger}] = \delta_{\beta\gamma} \sum_n\, \Lketc{\{ n_\alpha \}_A} \Lbrac{\{ n_\alpha \}_A}.
\end{equation}
Namely, the algebra between the canonical mirror operators having the same microstate index is exactly that of standard annihilation and creation operators.

\subsection{Global promotion}
\label{subsec:global}

The operators $\tilde{b}_\gamma^A$ and $\tilde{b}_\gamma^{A\dagger}$ described above are defined for each microstate $A$.
We can promote them to operators that act more ``globally'' in the space of microstates.
(For an analogous construction, see Ref.~\cite{Papadodimas:2015jra}.)
Consider the Hilbert space spanned by all the independent vacuum microstates (of a given mass $M$)
\begin{equation}
  {\cal M} = \Biggl\{ \sum_{A=1}^{e^{S_{\rm tot}}} a_A \ket{\Psi_{A,0}(M)} \,\Bigg|\, a_A \in \mathbb{C},\, \sum_{A=1}^{e^{S_{\rm tot}}} |a_A|^2 = 1 \Biggr\}.
\label{eq:cal-M}
\end{equation}
Consider a subspace of ${\cal M}$ spanned by $e^{S_{\rm eff}}$ independent microstates 
\begin{equation}
  \tilde{\cal M} = \Biggl\{ \sum_{A'=1}^{e^{S_{\rm eff}}} a_{A'} \ket{\Psi_{A',0}(M)} \,\Bigg|\, a_{A'} \in \mathbb{C},\, \sum_{A'=1}^{e^{S_{\rm eff}}} |a_{A'}|^2 = 1 \Biggr\},
\label{eq:tilde-cal-M}
\end{equation}
where
\begin{equation}
  S_{\rm eff} < S_{\rm bh}(M) + S_{\rm rad}.
\label{eq:S_eff-cond}
\end{equation}
By choosing the bases of ${\cal M}$ and $\tilde{\cal M}$ appropriately, we can take $\{ \ket{\Psi_{A',0}(M)} \}$ to be a subset of $\{ \ket{\Psi_{A,0}(M)} \}$, i.e.\ $A' \subset A$.
This leads to
\begin{equation}
  \inner{\Psi_{A',0}(M)}{\Psi_{B',0}(M)} = \delta_{A'B'}.
\end{equation}

We define {\it globally promoted canonical mirror operators} associated with Hilbert subspace $\tilde{\cal M}$ by
\begin{align}
  \tilde{\cal B}_\gamma &= \sum_{A'=1}^{e^{S_{\rm eff}}} \tilde{b}_\gamma^{A'},
\label{eq:global-ann}\\
  \tilde{\cal B}_\gamma^\dagger &= \sum_{A'=1}^{e^{S_{\rm eff}}} \tilde{b}_\gamma^{A'\dagger},
\label{eq:global-cre}
\end{align}
where $\tilde{b}_\gamma^{A'}$ and $\tilde{b}_\gamma^{A'\dagger}$ are given by Eqs.~(\ref{eq:ann-m-orig},~\ref{eq:cre-m-orig}).
The commutation relations of these operators are
\begin{align}
  [\tilde{\cal B}_\beta, \tilde{\cal B}_\gamma] &= \sum_{\substack{A',B'=1 \\ A' \neq B'}}^{e^{S_{\rm eff}}} \sum_n \sqrt{n_\beta n_\gamma}\,\, \zeta_{n_\smg n_\smb}^{A'B'} \Lketc{\{ n_\alpha - \delta_{\alpha\beta} - \delta_{\alpha\gamma} \}_{A'}} \Lbrac{\{ n_\alpha \}_{B'}},
\\
  [\tilde{\cal B}_\beta^\dagger, \tilde{\cal B}_\gamma^\dagger] &= \sum_{\substack{A',B'=1 \\ A' \neq B'}}^{e^{S_{\rm eff}}} \sum_n \sqrt{(n_\beta+1) (n_\gamma+1)}\,\, \zeta_{n_\spg n_\spb}^{A'B'} \Lketc{\{ n_\alpha + \delta_{\alpha\beta} + \delta_{\alpha\gamma} \}_{A'}} \Lbrac{\{ n_\alpha \}_{B'}},
\\
  [\tilde{\cal B}_\beta, \tilde{\cal B}_\gamma^\dagger] &= \begin{cases}
  \sum_{A'=1}^{e^{S_{\rm eff}}} \sum_n\, \Lketc{\{ n_\alpha \}_{A'}} \Lbrac{\{ n_\alpha \}_{A'}} & 
\nonumber\\*
  \qquad + \sum_{\substack{A',B'=1 \\ A' \neq B'}}^{e^{S_{\rm eff}}} \sum_n\, \bigl\{ (n_\gamma + 1)\varepsilon_{n_\spg}^{A'B'} - n_\gamma\varepsilon_{n_\smg}^{A'B'}) \bigr\}\, \Lketc{\{ n_\alpha \}_{A'}} \Lbrac{\{ n_\alpha \}_{B'}}
  \qquad \mbox{for } \beta = \gamma
\\
  \sum_{\substack{A',B'=1 \\ A' \neq B'}}^{e^{S_{\rm eff}}} \sum_n \sqrt{n_\beta (n_\gamma + 1)}\, (\varepsilon_{n_\spg}^{A'B'} - \varepsilon_{n_\smb}^{A'B'})\, \Lketc{\{ n_\alpha - \delta_{\alpha\beta} + \delta_{\alpha\gamma} \}_{A'}} \Lbrac{\{ n_\alpha \}_{B'}}
  \qquad \mbox{for } \beta \neq \gamma,
\end{cases}
\end{align}
where
\begin{align}
  \zeta_{n_\smg n_\smb}^{A'B'} \,\equiv\, \varepsilon_{n_\smg}^{A'B'} - \varepsilon_{n_\smb}^{A'B'} = O\left( \frac{e^{-\frac{\omega_\gamma}{2T_{\rm H}}} - e^{-\frac{\omega_\beta}{2T_{\rm H}}}}{e^{\frac{1}{2}S_{\rm bh}(M-E_n)+\frac{1}{2}S_{\rm rad}}} \right),
\\*
  \zeta_{n_\spg n_\spb}^{A'B'} \,\equiv\, \varepsilon_{n_\spg}^{A'B'} - \varepsilon_{n_\spb}^{A'B'} = O\left( \frac{e^{\frac{\omega_\gamma}{2T_{\rm H}}} - e^{\frac{\omega_\beta}{2T_{\rm H}}}}{e^{\frac{1}{2}S_{\rm bh}(M-E_n)+\frac{1}{2}S_{\rm rad}}} \right).
\end{align}

The matrix elements of the globally promoted operators between mirror microstates $\Lketc{\{ \kappa_\alpha \}_{E'}}$ and $\Lketc{\{ \lambda_\alpha \}_{F'}}$ are given by
\begin{align}
  & \Lbrac{\{ \kappa_\alpha \}_{E'}}\, \tilde{\cal B}_\gamma\, \Lketc{\{ \lambda_\alpha \}_{F'}} = \delta_{\kappa \lambda_\smg} \sqrt{\lambda_\gamma}\, \left( \sum_{A'=1}^{e^{S_{\rm eff}}} \eta_{\lambda_\smg}^{E'A'} \eta_\lambda^{A'F'} \right),
\label{eq:ME-1_1}\\*
  & \Lbrac{\{ \kappa_\alpha \}_{E'}}\, \tilde{\cal B}_\gamma^\dagger\, \Lketc{\{ \lambda_\alpha \}_{F'}} = \delta_{\kappa \lambda_\spg} \sqrt{\lambda_\gamma + 1}\, \left( \sum_{A'=1}^{e^{S_{\rm eff}}} \eta_{\lambda_\spg}^{E'A'} \eta_\lambda^{A'F'} \right)
\label{eq:ME-1_2}
\end{align}
for annihilation and creation operators, and
\begin{align}
  & \Lbrac{\{ \kappa_\alpha \}_{E'}}\, \tilde{\cal B}_\beta \tilde{\cal B}_\gamma\, \Lketc{\{ \lambda_\alpha \}_{F'}} 
  = \delta_{\kappa \lambda_\smbmg} \sqrt{(\lambda_\beta - \delta_{\beta\gamma}) \lambda_\gamma}\, \left( \sum_{A'=1}^{e^{S_{\rm eff}}} \sum_{B'=1}^{e^{S_{\rm eff}}} \eta_{\lambda_\smbmg}^{E'A'} \eta_{\lambda_\smg}^{A'B'} \eta_\lambda^{B'F'} \right),
\label{eq:ME-2_1}\\
  & \Lbrac{\{ \kappa_\alpha \}_{E'}}\, \tilde{\cal B}_\beta^\dagger \tilde{\cal B}_\gamma^\dagger\, \Lketc{\{ \lambda_\alpha \}_{F'}} 
  = \delta_{\kappa \lambda_\spbpg} \sqrt{(\lambda_\beta + 1 + \delta_{\beta\gamma})(\lambda_\gamma + 1)}\, \left( \sum_{A'=1}^{e^{S_{\rm eff}}} \sum_{B'=1}^{e^{S_{\rm eff}}} \eta_{\lambda_\spbpg}^{E'A'} \eta_{\lambda_\spg}^{A'B'} \eta_\lambda^{B'F'} \right),
\label{eq:ME-2_2}\\
  & \Lbrac{\{ \kappa_\alpha \}_{E'}}\, \tilde{\cal B}_\beta \tilde{\cal B}_\gamma^\dagger\, \Lketc{\{ \lambda_\alpha \}_{F'}} 
  = \delta_{\kappa \lambda_\smbpg} \sqrt{(\lambda_\beta + \delta_{\beta\gamma}) (\lambda_\gamma + 1)}\, \left( \sum_{A'=1}^{e^{S_{\rm eff}}} \sum_{B'=1}^{e^{S_{\rm eff}}} \eta_{\lambda_\smbpg}^{E'A'} \eta_{\lambda_\spg}^{A'B'} \eta_\lambda^{B'F'} \right),
\label{eq:ME-2_3}\\
  & \Lbrac{\{ \kappa_\alpha \}_{E'}}\, \tilde{\cal B}_\beta^\dagger \tilde{\cal B}_\gamma\, \Lketc{\{ \lambda_\alpha \}_{F'}} 
  = \delta_{\kappa \lambda_\spbmg} \sqrt{(\lambda_\beta + 1 - \delta_{\beta\gamma}) \lambda_\gamma}\, \left( \sum_{A'=1}^{e^{S_{\rm eff}}} \sum_{B'=1}^{e^{S_{\rm eff}}} \eta_{\lambda_\spbmg}^{E'A'} \eta_{\lambda_\smg}^{A'B'} \eta_\lambda^{B'F'} \right)
\label{eq:ME-2_4}
\end{align}
for products of two operators.
Here, $\eta_\lambda^{A'F'}$ is given by Eq.~(\ref{eq:eta}), $\lambda_{-\beta-\gamma} = \{ \lambda_\alpha - \delta_{\alpha\beta} - \delta_{\alpha\gamma} \}$, and so on.
Similarly, for products of three and more operators
\begin{align}
  & \Lbrac{\{ \kappa_\alpha \}_{E'}}\, \tilde{\cal B}_\beta \tilde{\cal B}_\gamma \tilde{\cal B}_\rho\, \Lketc{\{ \lambda_\alpha \}_{F'}} 
  = \delta_{\kappa \lambda_{\scalebox{0.4}{$-\beta\!-\!\gamma\!-\!\rho$}}} \sqrt{(\lambda_\beta - \delta_{\beta\gamma} - \delta_{\beta\rho}) (\lambda_\gamma - \delta_{\gamma\rho}) \lambda_\rho}
\nonumber\\*
  &\qquad\qquad\qquad\qquad\qquad\qquad\quad\times 
\left( \sum_{A'=1}^{e^{S_{\rm eff}}} \sum_{B'=1}^{e^{S_{\rm eff}}} \sum_{C'=1}^{e^{S_{\rm eff}}} \eta_{\lambda_{\scalebox{0.4}{$-\beta\!-\!\gamma\!-\!\rho$}}}^{E'A'} \eta_{\lambda_{\scalebox{0.4}{$-\gamma\!-\!\rho$}}}^{A'B'} \eta_{\lambda{\scalebox{0.4}{$-\rho$}}}^{B'C'} \eta_\lambda^{C'F'} \right),
\\*
  & \cdots.
\nonumber
\end{align}

As will become clearer later, the relevant quantities are the combinations of $\eta$'s that appear in these matrix elements.
With Eq.~(\ref{eq:S_eff-cond}), Eqs.~(\ref{eq:ME-1_1},~\ref{eq:ME-1_2}) give
\begin{equation}
  \sum_{A'=1}^{e^{S_{\rm eff}}} \eta_\kappa^{E'A'} \eta_\lambda^{A'F'} \Bigr|_{\kappa \neq \lambda} 
  = \begin{cases} 
    1 + \sum_{\subalign{& A'=1 \\& A' \neq E'}}^{e^{S_{\rm eff}}} \varepsilon_\kappa^{E'A'} \varepsilon_\lambda^{A'E'} 
  = 1 + O\left( \frac{e^{\frac{E_\kappa + E_\lambda}{2T_{\rm H}}+\frac{1}{2}S_{\rm eff}}}{e^{S_{\rm bh}(M)+S_{\rm rad}}} \right)
  & \mbox{for } E' = F'
\\
    \varepsilon_\kappa^{E'F'} + \varepsilon_\lambda^{E'F'} + \sum_{\subalign{& A'=1 \\& A' \neq E',F'}}^{e^{S_{\rm eff}}} \varepsilon_\kappa^{E'A'} \varepsilon_\lambda^{A'F'} 
  = O\left( \frac{e^{\frac{E_{\rm max}}{2T_{\rm H}}}}{e^{\frac{1}{2}S_{\rm bh}(M)+\frac{1}{2}S_{\rm rad}}} \right)
  & \mbox{for } E' \neq F',
  \end{cases}
\end{equation}
where $E_{\rm max} = {\rm max}\{ E_\kappa, E_\lambda \}$, so we find that this quantity is $\delta_{E'F'}$ up to corrections exponentially suppressed by $e^{-\frac{1}{2}\{ S_{\rm bh}(M) + S_{\rm rad}\} + \frac{E_{\rm max}}{2T_{\rm H}}}$.
For Eqs.~(\ref{eq:ME-2_1}~--~\ref{eq:ME-2_4}), we find
\begin{align}
  & \sum_{A'=1}^{e^{S_{\rm eff}}} \sum_{B'=1}^{e^{S_{\rm eff}}} \eta_\kappa^{E'A'} \eta_\mu^{A'B'} \eta_\lambda^{B'F'} \Bigr|_{\mu \neq \kappa, \lambda} 
  = \begin{cases} 
    1 + \delta_{\kappa\lambda}\, O\left( \frac{e^{\frac{E_{\rm max}}{2T_{\rm H}}+S_{\rm eff}}}{e^{S_{\rm bh}(M)+S_{\rm rad}}} \right) + O\left( \frac{e^{\frac{E_{\rm max}}{2T_{\rm H}}+\frac{1}{2}S_{\rm eff}}}{e^{S_{\rm bh}(M)+S_{\rm rad}}} \right)
  & \mbox{for } E' = F'
\\
    O\left( \frac{e^{\frac{E_{\rm max}}{2T_{\rm H}}}}{e^{\frac{1}{2}S_{\rm bh}(M)+\frac{1}{2}S_{\rm rad}}} \right)
  & \mbox{for } E' \neq F',
  \end{cases}
\end{align}
where $E_{\rm max} = {\rm max}\{ E_\kappa, E_\mu, E_\lambda \}$.
The deviation of this quantity from $\delta_{E'F'}$ is thus suppressed if $e^{-S_{\rm bh}(M) - S_{\rm rad} + S_{\rm eff} + \frac{E_{\rm max}}{2T_{\rm H}}}$ is small.
Given that $E_{\rm max}/T_{\rm H}$ is much smaller than ${\rm min}\{ S_{\rm bh}(M), S_{\rm rad} \}$ for the backreaction on the black hole geometry to be small (except possibly for the very early stage of the black hole evolution, when $S_{\rm rad} \ll S_{\rm bh}(M)$), this condition is satisfied when
\begin{equation}
  \frac{S_{\rm bh}(M) + S_{\rm rad} - S_{\rm eff}}{S_{\rm bh}(M) + S_{\rm rad}} \nll 1.
\end{equation}
Namely, the condition is satisfied unless the fractional difference between $S_{\rm eff}$ and $S_{\rm bh}(M) + S_{\rm rad}$ is much smaller than $1$.

A similar analysis reveals that the situation is the same for larger numbers of operators.
In particular,
\begin{equation}
  \sum_{A_1'=1}^{e^{S_{\rm eff}}} \sum_{A_2'=1}^{e^{S_{\rm eff}}} \cdots \sum_{A_{p-1}'=1}^{e^{S_{\rm eff}}} \eta_{\lambda_1}^{E'A_1'} \eta_{\lambda_2}^{A_1'A_2'} \cdots \eta_{\lambda_p}^{A_{p-1}'F'} 
  = \delta_{E'F'} + \begin{cases} 
    O\left( \frac{1}{e^{\frac{1}{2}\{ S_{\rm bh}(M)+S_{\rm rad} \}}} \right) 
  & \mbox{for } S_{\rm eff} < \frac{1}{2} \{ S_{\rm bh}(M)+S_{\rm rad} \}
\vspace{0.2cm}\\
    O\left( \frac{e^{S_{\rm eff}}}{e^{S_{\rm bh}(M)+S_{\rm rad}}} \right)
  & \mbox{for } S_{\rm eff} > \frac{1}{2} \{ S_{\rm bh}(M)+S_{\rm rad} \}
  \end{cases}
\label{eq:correction}
\end{equation}
for arbitrary $p$ and $\lambda_{1,2,\cdots,p}$.
This implies that the matrix elements of products of globally promoted operators $\tilde{\cal B}_\gamma$ and $\tilde{\cal B}_\gamma^\dagger$ between states in $\tilde{\cal M}$ are the same as the corresponding quantities in field theory on the two-sided black hole geometry, up to corrections of order
\begin{equation}
  \epsilon = {\rm max}\left\{ \frac{1}{e^{\frac{1}{2}\{ S_{\rm bh}(M)+S_{\rm rad} \}}}, \frac{e^{S_{\rm eff}}}{e^{S_{\rm bh}(M)+S_{\rm rad}}} \right\}.
\label{eq:epsilon}
\end{equation}

\subsection{Infalling mode operators}
\label{subsec:infalling}

Using the operators defined so far, we can construct {\it infalling mode operators}, as described in Ref.~\cite{Papadodimas:2012aq}.
For each microstate $A$, we define
\begin{align}
  a_\xi^A &= \sum_\gamma \bigl( \alpha_{\xi\gamma} b_\gamma + \beta_{\xi\gamma} b_\gamma^\dagger + \zeta_{\xi\gamma} \tilde{b}_\gamma^A + \eta_{\xi\gamma} \tilde{b}_\gamma^{A\dagger} \bigr),
\label{eq:a_xi}\\*
  a_\xi^{A\dagger} &= \sum_\gamma \bigl( \beta_{\xi\gamma}^* b_\gamma + \alpha_{\xi\gamma}^* b_\gamma^\dagger + \eta_{\xi\gamma}^* \tilde{b}_\gamma^A + \zeta_{\xi\gamma}^* \tilde{b}_\gamma^{A\dagger} \bigr),
\label{eq:a_xi-dag}
\end{align}
where $\xi$ is the label in which the frequency $\omega$ with respect to boundary time $t$ is traded with the frequency $\Omega$ associated with infalling time $\tau$, and $\alpha_{\xi\gamma}$, $\beta_{\xi\gamma}$, $\zeta_{\xi\gamma}$, and $\eta_{\xi\gamma}$ are the Bogoliubov coefficients calculable using the standard field theory method.%
\footnote{For a massless scalar field, for example, Eq.~(\ref{eq:a_xi}) takes the form
 \begin{equation}
   a_\xi^A = \pm \frac{i}{2\pi \sqrt{\Omega_\xi T_{\rm H}}} \int_0^\infty\! d\omega_\gamma \left[ \frac{\Xi}{\sqrt{1 - e^{-\frac{\omega_\gamma}{T_{\rm H}}}}}\, b_\gamma + \frac{\Xi^*}{\sqrt{e^{\frac{\omega_\gamma}{T_{\rm H}}} - 1}}\, b_\gamma^\dagger - \frac{\Xi^*}{\sqrt{1 - e^{-\frac{\omega_\gamma}{T_{\rm H}}}}}\, \tilde{b}_\gamma^A - \frac{\Xi}{\sqrt{e^{\frac{\omega_\gamma}{T_{\rm H}}} - 1}}\, \tilde{b}_\gamma^{A\dagger} \right]
 \nonumber
 \end{equation}
 in the near horizon limit.
 Here, we have adopted the continuum notation for the sum over the frequency, and $\Xi = \left(\Omega_\xi/2\pi T_{\rm H}\right)^{\pm\frac{i\omega_\gamma}{2\pi T_{\rm H}}} \Gamma\left(1 \pm \frac{\omega_\gamma}{2\pi i T_{\rm H}}\right)/\left|\Gamma\left(1 \pm \frac{\omega_\gamma}{2\pi i T_{\rm H}}\right)\right|$ is a pure phase.
 The $\pm$ symbol in these equations takes $+$ and $-$ for ingoing and outgoing modes, respectively.}

As in Section~\ref{subsec:global}, we can also define globally promoted infalling mode operators:
\begin{align}
  {\cal A}_\xi &= \sum_\gamma \bigl( \alpha_{\xi\gamma} b_\gamma + \beta_{\xi\gamma} b_\gamma^\dagger + \zeta_{\xi\gamma} \tilde{\cal B}_\gamma + \eta_{\xi\gamma} \tilde{\cal B}_\gamma^\dagger \bigr),
\label{eq:A_xi}\\*
  {\cal A}_\xi^\dagger &= \sum_\gamma \bigl( \beta_{\xi\gamma}^* b_\gamma + \alpha_{\xi\gamma}^* b_\gamma^\dagger + \eta_{\xi\gamma}^* \tilde{\cal B}_\gamma + \zeta_{\xi\gamma}^* \tilde{\cal B}_\gamma^\dagger \bigr),
\label{eq:A_xi-dag}
\end{align}
which act linearly in the space $\tilde{\cal M}$ in Eq.~(\ref{eq:tilde-cal-M}).
From the results in Section~\ref{subsec:global}, we find that the deviations of the matrix elements of products of these operators between states in $\tilde{\cal M}$ from the corresponding field theory quantities on the two-sided geometry involve factors
\begin{equation}
  \sum_{A_1'=1}^{e^{S_{\rm eff}}} \sum_{A_2'=1}^{e^{S_{\rm eff}}} \cdots \sum_{A_{p-1}'=1}^{e^{S_{\rm eff}}} \varepsilon_{\lambda_1}^{E'A_1'} \varepsilon_{\lambda_2}^{A_1'A_2'} \cdots \varepsilon_{\lambda_p}^{A_{p-1}'F'}
\qquad
  (p = 1,2,\cdots)
\label{eq:factors}
\end{equation}
multiplied by functions of occupation numbers and Bogoliubov coefficients, integrated over the frequencies.
Here, $p=1$ in Eq.~(\ref{eq:factors}) means $\varepsilon_{\lambda_1}^{E'F'}$.
An analysis similar to that in the previous subsection shows that these quantities are suppressed exponentially by the larger of $e^{-\frac{1}{2}\{S_{\rm bh}(M)+S_{\rm rad}\}}$ and $e^{S_{\rm eff} - \{S_{\rm bh}(M)+S_{\rm rad}\}}$.
Namely, the matrix elements of products of ${\cal A}_\xi$ and ${\cal A}_\xi^\dagger$ in $\tilde{\cal M}$ are the same as the corresponding field theory values, up to corrections suppressed by $\epsilon$ in Eq.~(\ref{eq:epsilon}).

\section{Effective Theory of the Interior}
\label{sec:eff-theory}

In this section, we detail how an effective theory describing the black hole interior can be erected using operators described in Section~\ref{sec:operators}.
We first present the minimal construction of the theory and discuss an ambiguity existing in the construction.
This ambiguity is viewed as an intrinsic ambiguity of a semiclassical description.
We then discuss the (in)dependence of infalling operators on the microstate of the system and a connection of the resulting picture to quantum error correction.
We also comment on the situation for a young black hole and the Minkowski limit.

\subsection{Emergence of a stable semiclassical description}
\label{subsec:eff-space}

Suppose that the state of the system at a boundary time $t_*$ is given by Eq.~(\ref{eq:Psi-t_gen}).
Since our interest is the black hole in question, for the excitations labeled by $I$ we disregard those outside the zone (the inclusion of which is straightforward).
As before, we focus on a branch in which the mass of the black hole is $M$ (in the sense discussed around Eq.~(\ref{eq:Psi-t_gen})), and we adopt the notation in Eq.~(\ref{eq:A_M-A}).
Our state is thus
\begin{equation}
  \ket{\Psi(t_*)} = \sum_{A=1}^{S_{\rm tot}} \sum_I d_{A I}(t_*) \ket{\Psi_{A,I}(M)},
\label{eq:Psi-t*}
\end{equation}
where $I$ labels excitations in the zone.%
\footnote{These excitations include objects in the interior region in the effective theory if the black hole state (after removing semiclassical objects in the zone) is not yet equilibrated.
 This will be discussed in more detail later.}

We want to understand what an object located in the zone and falling toward the black hole will experience after it crosses the horizon.
For this purpose, a description based on boundary time evolution is of little use.
In that description, the object will be absorbed into the stretched horizon when it gets there, after which no low energy description is available for it.
To describe the object's experience after reaching the horizon, we need a different time evolution associated with the proper time of the object.

In general, the state in Eq.~(\ref{eq:Psi-t*}) does not factorize into a product of two states having indices $A$ and $I$, respectively, since the degrees of freedom represented by the two indices can be entangled due to their interactions in the past, for example through interactions before the black hole is formed or through interactions between the object and Hawking radiation.
One might still think that since we can expand the state in terms of a basis in $A$ space as
\begin{equation}
  \ket{\Psi(t_*)} = \sum_{A=1}^{e^{S_{\rm tot}}} f_A \ket{\Psi_{A,I(A)}(M)},
\label{eq:Psi-t*_exp}
\end{equation}
where $\sum_{A=1}^{e^{S_{\rm tot}}} |f_A|^2 = 1$, we can regard it as representing $e^{S_{\rm tot}}$ decohered branch worlds and describe the future evolution of excitations for each of them using the infalling mode operators in Eqs.~(\ref{eq:a_xi},~\ref{eq:a_xi-dag}).%
\footnote{This corresponds to the picture presented in Refs.~\cite{Nomura:2012ex,Bao:2017who}.}
This would indeed be sufficient if the ``observer''---a classical system that is separated from the measured system as an external structure---lives purely in a single such microbranch.
This, however, is not the case in general.

If the observer is not restricted to a single microbranch, we must deal with the density matrix in $I$ space, which takes the form
\begin{equation}
  \rho(t_*) = \sum_{A=1}^{e^{S_{\rm tot}}} |f_A|^2 \ket{\Psi_{A,I(A)}(M)} \bra{\Psi_{A,I(A)}(M)}.
\label{eq:rho-t*}
\end{equation}
In this case, we would a priori have to use different infalling mode operators for different terms, i.e.\ Eqs.~(\ref{eq:a_xi},~\ref{eq:a_xi-dag}) with $A$ taking the value corresponding to each term.
This is somewhat uncomfortable.
Moreover, the infalling mode operators that must be used even depend on the basis of $A$ one takes in writing Eq.~(\ref{eq:Psi-t*_exp}).
Specifically, if we choose to expand the state in Eq.~(\ref{eq:Psi-t*_exp}) in terms of another set of basis states labeled by $A'$, then we get
\begin{equation}
  \rho(t_*) = \sum_{A'=1}^{e^{S_{\rm tot}}} |f_{A'}|^2 \ket{\Psi_{A',I(A')}(M)} \bra{\Psi_{A',I(A')}(M)}.
\label{eq:rho-t*-2}
\end{equation}
This is, of course, the same reduced density matrix as Eq.~(\ref{eq:rho-t*}) in $I$ space.
Nevertheless, the infalling mode operators used in each term are now Eqs.~(\ref{eq:a_xi},~\ref{eq:a_xi-dag}) with $A$ replaced with $A'$; i.e., predicting the outcome of a measurement performed in a subsystem requires knowledge beyond the reduced density matrix of the subsystem.
This clearly violates the principles of standard quantum mechanics.

The existence of globally promoted operators, however, provides a rescue.
(For related discussion, see Ref.~\cite{Papadodimas:2015jra}.)
Suppose that the state at boundary time $t_*$ is given by Eq.~(\ref{eq:Psi-t*}).
Given that the dimension of the excitation Hilbert space is much smaller than $e^{S_{\rm tot}}$, we can write it using the Schmidt decomposition as
\begin{equation}
  \ket{\Psi(t_*)} = \sum_{I=1}^{\cal K} g_I \ket{\Psi_{A(I),I}(M)},
\label{eq:Psi-t*_Sch}
\end{equation}
where $\sum_{I=1}^{\cal K} |g_I|^2 = 1$, $g_I > 0$, and ${\cal K}$ is the Schmidt number.
The point is that ${\cal K}$ always satisfies
\begin{equation}
  {\cal K} \leq S_{\rm exc} < S_{\rm bh}(M) + S_{\rm rad},
\end{equation}
where $S_{\rm exc}$ is the logarithm of the dimension of the Hilbert space for semiclassical excitations (hard modes).
Therefore, by taking the observable to be the globally promoted infalling mode operators, ${\cal A}_\xi$ and ${\cal A}_\xi^\dagger$, with $\tilde{\cal M}$ containing
\begin{equation}
  \tilde{V}\bigl[\ket{\Psi(t_*)}\bigr] = {\rm span}\bigl( \{ \ket{\Psi_{A(I),0}(M)} \} \bigr),
\end{equation}
i.e.
\begin{equation}
  \tilde{\cal M} \supseteq \tilde{V}\bigl[\ket{\Psi(t_*)}\bigr],
\label{eq:tilde-M_1}
\end{equation}
we can preserve the tenets of quantum mechanics.%
\footnote{A more precise condition for the choice of $\tilde{\cal M}$ will be discussed in Section~\ref{subsec:state-dep}.}
Note that the dimension of $\tilde{\cal M}$ can be anything that satisfies
\begin{equation}
  {\cal K} = \ln{\rm dim}\,\tilde{V}\bigl[\ket{\Psi(t_*)}\bigr] \,\leq\, \ln{\rm dim}\,\tilde{\cal M} \,<\, S_{\rm bh}(M) + S_{\rm rad},
\end{equation}
unless the fractional difference between $\ln{\rm dim}\,\tilde{\cal M}$ and $S_{\rm bh}(M) + S_{\rm rad}$ is exponentially small.
Since
\begin{equation}
  S_{\rm bh}(M) + S_{\rm rad} \,<\, S_{\rm tot} = \ln{\rm dim}\,{\cal M},
\end{equation}
$\tilde{\cal M}$ is a proper subset of the microscopic vacuum Hilbert space ${\cal M}$ in Eq.~(\ref{eq:cal-M}).

With the infalling mode operators chosen in this way, we can construct the infalling Hamiltonian
\begin{equation}
  \tilde{H} = \sum_\xi \Omega_\xi {\cal A}_\xi^\dagger {\cal A}_\xi + \tilde{H}_{\rm int}\bigl( \{ {\cal A}_\xi \}, \{ {\cal A}_\xi^\dagger \} \bigr),
\label{eq:tilde-H}
\end{equation}
where $\tilde{H}_{\rm int}(\{ {\cal A}_\xi \}, \{ {\cal A}_\xi^\dagger \})$ is determined by matching it with $H_{\rm int}(\{ b_\gamma \}, \{ b_\gamma^\dagger \})$ in Eq.~(\ref{eq:H}) in the first exterior.
The explicit form of $\tilde{H}_{\rm int}(\{ {\cal A}_\xi \}, \{ {\cal A}_\xi^\dagger \})$ depends on the time parametrization we take, which is reflected in the Bogoliubov coefficients in Eqs.~(\ref{eq:A_xi},~\ref{eq:A_xi-dag}).

\subsection{Erecting an effective theory}
\label{subsec:match}

The initial state of the infalling time evolution is given by mapping the states on the right-hand side of Eq.~(\ref{eq:Psi-t*_Sch}) to those in the effective two-sided geometry:
\begin{equation}
  \ket{\Psi_{A(I),I}(M)} \,\mapsto\, \ket{\tilde{\Psi}_{A(I),I}},
\label{eq:convert}
\end{equation}
where the states in the two-sided geometry, $\ket{\tilde{\Psi}_{A(I),I}}$, are defined on the union $U_0$ of the zone and its mirror region on the hypersurface of infalling time $\tau = 0$, which we match with the boundary time $t_*$ (see Fig.~\ref{fig:two-sided}).
Note that this operation is not tracing out the far region in the original one-sided theory; rather, it is a map of states of the original theory into those in the effective two-sided theory defined on the finite spatial region $U_0$.

The index $I$ of $\ket{\Psi_{A(I),I}(M)}$ labels excitations over the semiclassical black hole vacuum, and $A(I)$ specifies the corresponding vacuum microstate.
There are two types of an excitation represented by $I$.
One is an excitation of hard modes corresponding to an object in the zone, and the other is a (sufficiently large) deviation of the black hole state from the equilibrium form of Eq.~(\ref{eq:sys-state}) which cannot be attributed only to the hard modes.
Suppose that a falling object hits the stretched horizon at $t = t_*$.
The object then disappears from the zone, but it does not mean that the state immediately becomes the equilibrium form of Eq.~(\ref{eq:sys-state}) at $t = t_*$; rather, it stays in an excited form deviating from Eq.~(\ref{eq:sys-state}) for a while.
Now, consider that we erect an effective theory of the interior shortly after $t_*$:\ $t = t_* + \delta t$.
In this case, the effective theory is expected to have semiclassical excitations in the interior reflecting the fact that there is an object that has fallen into the horizon at $t_*$.
These include everything that the object does to the spacetime region described by the effective theory; for example, the object may emit a high energy quantum in the outward direction shortly after it crosses the horizon.
What the effective theory finds in the interior, represented by the index $I$ of $\ket{\tilde{\Psi}_{A(I),I}}$ on the right-hand side of Eq.~(\ref{eq:convert}), must be the excitations consistent with this semiclassical expectation, which is possible.%
\footnote{In the distant description, these excitations can be viewed as those of the stretched horizon, where the soft modes are mostly located.
 At the classical level, they correspond to an object that reaches $r = r_{\rm s}$ between $t = t_* - t_{\rm scr}$ and $t_*$ and is squeezed into the region between $r = r_+$ and $r_{\rm s}$ at time $t_*$ because of a large Lorentz contraction, where $r_+$ is the horizon radius.
 Understanding the map between the collective excitations of the soft modes, included in $I$ of $\ket{\Psi_{A(I),I}(M)}$, and these kinematically squeezed excitations in the effective theory, included in $I$ of $\ket{\tilde{\Psi}_{A(I),I}}$, would require a knowledge of the fundamental theory.}
This provides a built-in mechanism in the framework which allows for evading the ``frozen vacuum'' problem of Ref.~\cite{Bousso:2013ifa}.

On the other hand, if we erect an effective theory more than the scrambling time after the last disturbance to the stretched horizon, $t = t_* + \varDelta t$ with $\varDelta t > t_{\rm scr}$, then we expect that the effective theory finds the semiclassical vacuum in the interior, since the black hole state is already equilibrated by then (except for possible hard mode excitations corresponding to semiclassical objects in the zone).
This is indeed consistent with the semiclassical expectation as we see in more detail in the appendix.
In this case, the map of excitations in Eq.~(\ref{eq:convert}) is relatively straightforward; we simply have to convert excitations that are generated by $b_\gamma$ and $b_\gamma^\dagger$ and localized in the zone into those by ${\cal A}_\xi$ and ${\cal A}_\xi^\dagger$ using the semiclassical relation in Eqs.~(\ref{eq:A_xi},~\ref{eq:A_xi-dag}).

With the matching of states described above, the infalling time evolution can now be performed using the infalling Hamiltonian $\tilde{H}$ in Eq.~(\ref{eq:tilde-H}) with the initial state given by
\begin{equation}
  \tilde{\rho}(0) = \sum_{I=1}^{\cal K} |g_I|^2 \ket{\tilde{\Psi}_{A(I),I}} \bra{\tilde{\Psi}_{A(I),I}}.
\label{eq:rho-0}
\end{equation}
Since this state is defined only on the finite spatial region $U_0$, we need boundary conditions for the time evolution.
The physics we predict in the domain of dependence, $D(U_0)$, of $U_0$, however, does not depend on these boundary conditions.%
\footnote{An alternative, perhaps more physical, way to give the initial state is to specify it on a hypersurface consisting of the $\tau = 0$ surface in the zone of the first exterior and the intersection between the future horizon of the second exterior and $D(U_0)$ (properly smoothed to become a spacelike hypersurface).
 The initial state in Eq.~(\ref{eq:rho-0}) is then obtained by evolving this state backward in time using the Hamiltonian in Eq.~(\ref{eq:tilde-H}) (properly adapted to the new equal-time hypersurfaces) with a boundary condition chosen on the future boundary of $D(U_0)$ in the second exterior.
 Note that this implies that the ``map'' in Eq.~(\ref{eq:convert}) is not really a map in the mathematical sense; there are multiple choices of $\ket{\tilde{\Psi}_{A(I),I}}$ for the same $\ket{\Psi_{A(I),I}(M)}$ due to the freedom in this boundary condition, which all lead to the same physics in the interior of the black hole (but not in the fictitious, second exterior).}
(To cover a larger portion of the black hole interior, we must use multiple effective theories erected at different boundary times~\cite{Nomura:2018kia,Nomura:2019qps}.)

We stress that the theory of the interior obtained in this way is intrinsically semiclassical.
The information about the microstate is already traced out, which is reflected in the fact that the initial state in Eq.~(\ref{eq:rho-0}) is generally mixed.
Any future interactions between semiclassical and microscopic degrees of freedom can be treated only statistically, analogous to Hawking radiation in semiclassical calculations.
The theory is also not fully unitary under the time evolution by Eq.~(\ref{eq:tilde-H}); in particular, there is a singularity at $r = 0$ that cannot be resolved.

\subsection{Intrinsic ambiguity}
\label{subsec:ambiguity}

The construction of the effective theory described above has an ambiguity coming from the fact that the actions of infalling mode operators are not strictly orthogonal to $\tilde{\cal M}$ in the space of microstates.
We can see this, for example, by using Eqs.~(\ref{eq:non-ortho-1},~\ref{eq:non-ortho-2}) and Eqs.~(\ref{eq:ME-2_1}~--~\ref{eq:ME-2_4}):
\begin{align}
  \bra{\Psi_{A,0}(M)} b_\gamma^\dagger b_\gamma \ket{\Psi_{B,0}(M)} 
  &= \frac{1}{z} \sum_n n_\gamma\, e^{-\frac{E_n}{T_{\rm H}}} \left(\delta_{AB} + \varepsilon_n^{AB}\right)
\nonumber\\*
  &= \vev{n_\gamma}_{\rm ft} \delta_{AB} + O\left( \frac{1}{e^{\frac{1}{2}\{ S_{\rm bh}(M)+S_{\rm rad} \}}} \right)
,
\\*
  \bra{\Psi_{A,0}(M)} b_\gamma b_\gamma^\dagger \ket{\Psi_{B,0}(M)} 
  &= \frac{1}{z} \sum_n (n_\gamma + 1)\, e^{-\frac{E_n}{T_{\rm H}}} \left(\delta_{AB} + \varepsilon_n^{AB}\right)
\nonumber\\*
  &= \vev{n_\gamma + 1}_{\rm ft} \delta_{AB} + O\left( \frac{1}{e^{\frac{1}{2}\{ S_{\rm bh}(M)+S_{\rm rad} \}}} \right),
\end{align}
where $\vev{\cdots}_{\rm ft}$ represents the corresponding field theory values, and
\begin{align}
  \Lbrac{\{ \kappa_\alpha \}_{A'}}\, \tilde{\cal B}_\beta \tilde{\cal B}_\gamma\, \Lketc{\{ \lambda_\alpha \}_{B'}} 
  &= \delta_{\kappa \lambda_\smbmg} \sqrt{(\lambda_\beta - \delta_{\beta\gamma}) \lambda_\gamma}\, \left\{ \delta_{A'B'} + O\left( \frac{1}{e^{\frac{1}{2}\{ S_{\rm bh}(M)+S_{\rm rad} \}}} \right) \right\},
\\
  \Lbrac{\{ \kappa_\alpha \}_{A'}}\, \tilde{\cal B}_\beta^\dagger \tilde{\cal B}_\gamma^\dagger\, \Lketc{\{ \lambda_\alpha \}_{B'}} 
  &= \delta_{\kappa \lambda_\spbpg} \sqrt{(\lambda_\beta + 1 + \delta_{\beta\gamma})(\lambda_\gamma + 1)}\, \left\{ \delta_{A'B'} + O\left( \frac{1}{e^{\frac{1}{2}\{ S_{\rm bh}(M)+S_{\rm rad} \}}} \right) \right\},
\\
  \Lbrac{\{ \kappa_\alpha \}_{A'}}\, \tilde{\cal B}_\beta \tilde{\cal B}_\gamma^\dagger\, \Lketc{\{ \lambda_\alpha \}_{B'}} 
  &= \delta_{\kappa \lambda_\smbpg} \sqrt{(\lambda_\beta + \delta_{\beta\gamma}) (\lambda_\gamma + 1)}
\nonumber\\*
  & \quad \times \left\{ \delta_{A'B'} + \delta_{\beta\gamma}\,  O\left( \frac{e^{S_{\rm eff}}}{e^{S_{\rm bh}(M)+S_{\rm rad}}} \right) + O\left( \frac{1}{e^{\frac{1}{2}\{ S_{\rm bh}(M)+S_{\rm rad} \}}} \right) \right\},
\label{eq:O-qual-1}\\
  \Lbrac{\{ \kappa_\alpha \}_{A'}}\, \tilde{\cal B}_\beta^\dagger \tilde{\cal B}_\gamma\, \Lketc{\{ \lambda_\alpha \}_{B'}} 
  &= \delta_{\kappa \lambda_\spbmg} \sqrt{(\lambda_\beta + 1 - \delta_{\beta\gamma}) \lambda_\gamma}
\nonumber\\*
  & \quad \times \left\{ \delta_{A'B'} + \delta_{\beta\gamma}\, O\left( \frac{e^{S_{\rm eff}}}{e^{S_{\rm bh}(M)+S_{\rm rad}}} \right) + O\left( \frac{1}{e^{\frac{1}{2}\{ S_{\rm bh}(M)+S_{\rm rad} \}}} \right) \right\},
\label{eq:O-qual-2}
\end{align}
where the last $O(X)$ in Eqs.~(\ref{eq:O-qual-1},~\ref{eq:O-qual-2}) means that the corrections are smaller than $O(X)$ and does not necessarily imply the existence of corrections at that order.
The existence of exponentially suppressed corrections in these expressions---which are of order $\epsilon$ in Eq.~(\ref{eq:epsilon})---means that we cannot associate an excited state with a unique vacuum microstate in a strict sense.

The fact that the corrections are only of order $\epsilon$, however, implies that up to these exponentially suppressed corrections, the mode operators $b_\gamma$, $b_\gamma^\dagger$, $\tilde{\cal B}_\gamma$, $\tilde{\cal B}_\gamma^\dagger$, ${\cal A}_\xi$, and ${\cal A}_\xi^\dagger$ act only on the excitation index $I$, and not on the vacuum index $A'$.
In other words, ignoring these corrections, the Hilbert space can be viewed as
\begin{equation}
  {\cal H} \approx {\cal H}_{\rm exc} \otimes ({\cal H}_{\rm vac} \cong \tilde{\cal M}),
\label{eq:prod}
\end{equation}
where these mode operators act only on ${\cal H}_{\rm exc}$, and an excited state can be associated ``uniquely'' with a vacuum microstate.

The exponentially suppressed corrections discussed here constitute an intrinsic ambiguity of semiclassical physics, i.e.\ physics associated with the operators $b_\gamma$ and $b_\gamma^\dagger$ in a distant frame and ${\cal A}_\xi$ and ${\cal A}_\xi^\dagger$ in an infalling frame.
The maximal precision allowed by this ambiguity can be achieved by taking $\tilde{\cal M}$ as small as possible, i.e.
\begin{equation}
  \tilde{\cal M} = \tilde{V}\bigl[\ket{\Psi(t_*)}\bigr].
\label{eq:tilde-M_2}
\end{equation}
With this choice, the errors of the theory are of order $\epsilon$ in Eq.~(\ref{eq:epsilon}).

\subsection{Interior correlators in the in-in formalism}
\label{subsec:correlators}

We have seen that the matching of the states between the original (one-sided) and effective (two-sided) theories is given by Eq.~(\ref{eq:convert}), leading to the initial state for the infalling time evolution in Eq.~(\ref{eq:rho-0}).
In the standard practice in quantum field theory, this time evolution is implemented as that of quantum field operators in the Heisenberg picture.
How do we do this explicitly?

The operators at $\tau = 0$ are matched to those of the original theory at $t = t_*$ as Eqs.~(\ref{eq:A_xi},~\ref{eq:A_xi-dag}).
Quantum field operators at $\tau = 0$ are then given by
\begin{equation}
  \tilde{\Phi}_a({\bf x},0) = \sum_{s,\Omega,{\bf L}} \left( {\cal A}_\xi\, f_s(\Omega,{\bf L})\, \varphi_{\Omega,{\bf L}}({\bf x}) + {\cal A}_{\xi^c}^\dagger\, g_s(\Omega,{\bf L})\, \varphi_{\Omega,{\bf L}}^*({\bf x}) \right),
\end{equation}
where we have decomposed index $\xi$ into $a$, $s$, $\Omega$, and ${\bf L}$ which represent species, spin, frequency, and orbital angular momentum quantum numbers, respectively:\ $\xi = \{ a, s, \Omega, {\bf L} \}$.
Here, $f_s(\Omega,{\bf L})$ and $g_s(\Omega,{\bf L})$ are the standard factors providing Lorentz representation of the field (Dirac spinors, polarization vectors, etc), and $\varphi_{\Omega,{\bf L}}({\bf x})$ are the spatial wavefunctions.

The Heisenberg picture field operators are given by
\begin{equation}
  \tilde{\Phi}_a({\bf x},\tau) = e^{i\tilde{H}\tau} \tilde{\Phi}_a({\bf x},0) e^{-i\tilde{H}\tau},
\end{equation}
where $\tilde{H}$ is given in Eq.~(\ref{eq:tilde-H}).
The quantities we are interested in are correlators
\begin{equation}
  \left\langle \tilde{\Phi}_{a_1}(x_1) \tilde{\Phi}_{a_2}(x_2) \,\cdots\, \tilde{\Phi}_{a_n}(x_n) \right\rangle = {\rm Tr}\left[ \tilde{\rho}(0) \tilde{\Phi}_{a_1}(x_1) \tilde{\Phi}_{a_2}(x_2) \,\cdots\, \tilde{\Phi}_{a_n}(x_n) \right],
\label{eq:correlators}
\end{equation}
where $x_i = \{ {\bf x}_i, \tau_i \}$, and $\tilde{\rho}(0)$ is given by Eq.~(\ref{eq:rho-0}).
Since these are expectation values in the state given at a fixed finite time, $\tau = 0$, we must adopt the in-in formalism rather than the more conventional in-out formalism to calculate them.
This ultimately comes from the fact that the $S$-matrix cannot be defined at the semiclassical level for an object falling into a black hole.

\begin{figure}[t]
\begin{center}
  \includegraphics[height=3cm]{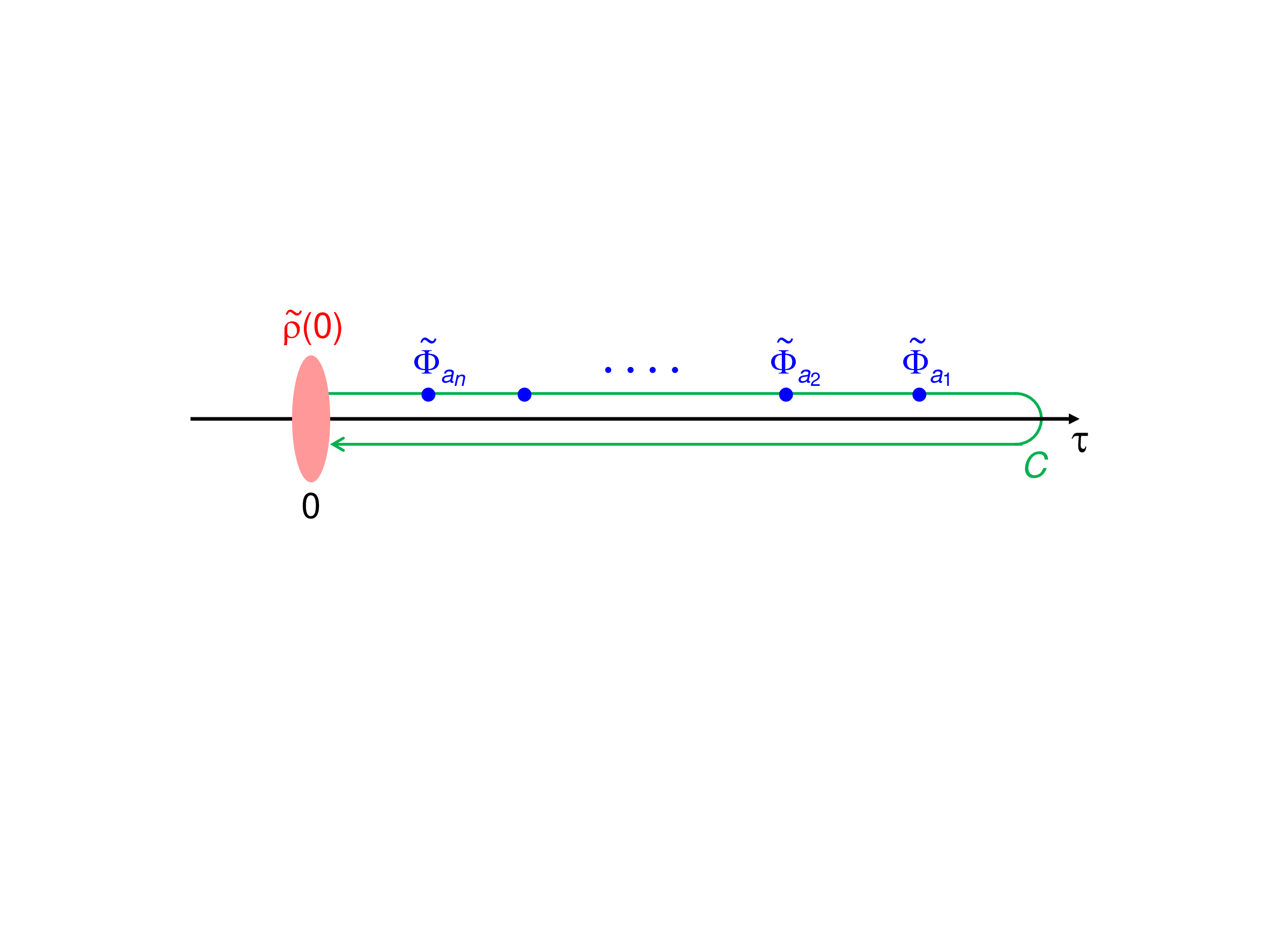}
\end{center}
\caption{The contour of integration $C$ for a time-ordered $n$-point correlator $\left\langle T\{\tilde{\Phi}_{a_1}(x_1)\, \tilde{\Phi}_{a_2}(x_2)\, \cdots\, \tilde{\Phi}_{a_n}(x_n)\} \right\rangle$ in the Schwinger-Keldysh formalism.
 Here, we have assumed $x_n^0 < \cdots < x_2^0 < x_1^0$ for illustration purposes.}
\label{fig:contour}
\end{figure}

Using the Schwinger-Keldysh method, a correlator of the form of Eq.~(\ref{eq:correlators}) can be written as a path integral over an appropriate closed time contour with the boundary condition given by $\tilde{\rho}(0)$.
For example, a time-ordered $n$-point correlator is given by taking the contour in Fig.~\ref{fig:contour}, yielding
\begin{equation}
  \left\langle T\{\tilde{\Phi}_{a_1}(x_1)\, \tilde{\Phi}_{a_2}(x_2)\, \cdots\, \tilde{\Phi}_{a_n}(x_n)\} \right\rangle = \sum_{I=1}^{\cal K} |g_I|^2\! \int\! \biggl( \prod_a {\cal D}\tilde{\Phi}^+_a {\cal D}\tilde{\Phi}^-_a \biggr)_{\!I}\! \tilde{\Phi}^+_{a_1}(x_1)\, \tilde{\Phi}^+_{a_2}(x_2)\, \cdots\, \tilde{\Phi}^+_{a_n}(x_n)\, e^{i S_{\rm SK}}.
\end{equation}
Here, $\int\! (\prod_a\! {\cal D}\tilde{\Phi}^+_a {\cal D}\tilde{\Phi}^-_a)_I$ represents path integral along the contour with the boundary conditions at $\tau = 0$ determined by $\ket{\tilde{\Psi}_{A(I),I}}$, and
\begin{equation}
  S_{\rm SK} = S\bigl[ \tilde{\Phi}^+_a \bigr] - S\bigl[ \tilde{\Phi}^-_a \bigr],
\end{equation}
where $S[\tilde{\Phi}_a]$ is the infalling frame action corresponding to $\tilde{H}$.
With this formalism, we can calculate arbitrary correlators as long as the fields are inside the domain of dependence of $U_0$.
Note that the field need not be in the interior of the black hole, so we can compute correlators between fields inside and outside the horizon.

As in standard quantum field theory, we can develop a perturbation theory to calculate the correlators.
For this purpose, we consider interaction picture fields
\begin{equation}
  \tilde{\phi}_a({\bf x},\tau) = e^{i\tilde{H}_0 \tau} \tilde{\Phi}_a({\bf x},0) e^{-i\tilde{H}_0 \tau},
\end{equation}
where $\tilde{H}_0$ is the free Hamiltonian, and regard ${\cal A}_\xi$ and ${\cal A}_\xi^\dagger$ as mode operators associated with them.
We also write $\ket{\tilde{\Psi}_{A(I),I}}$ in the initial state $\tilde{\rho}(0)$ in Eq.~(\ref{eq:rho-0}) as
\begin{equation}
  \ket{\tilde{\Psi}_{A(I),I}} = {\cal F}_I\bigl(\{ {\cal A}_\xi^\dagger \}\bigr)\, \ket{0},
\label{eq:Psi-inf-init}
\end{equation}
where $\ket{0}$ is the infalling vacuum defined by $\forall \xi, {\cal A}_\xi \ket{0} = 0$.
We can then use the canonical in-in formalism to calculate the correlators perturbatively.

\subsection{State dependence and quantum error correction}
\label{subsec:state-dep}

We have seen that in the existence of a black hole, there are a set of operators ${\cal A}_\xi$ and ${\cal A}_\xi^\dagger$ which act on hard, soft, and far modes in such a way that they are semiclassical annihilation and creation operators in the two-sided black hole background.
These operators have been chosen in a state dependent manner, but how sensitively do they depend on the microstate of the system?

\begin{figure}[t]
\begin{center}
  \includegraphics[height=6cm]{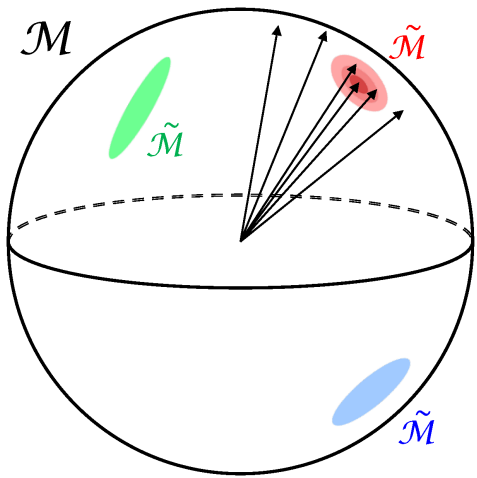}
\end{center}
\caption{In the $e^{S_{\rm tot}}$-dimensional space ${\cal M}$ spanned by orthonormal vacuum microstates, we can build infalling mode operators that cover a subspace $\tilde{\cal M}$ in a state independent manner.
 The choice of $\tilde{\cal M}$ is arbitrary as represented by regions with different colors.
 Furthermore, the space $\tilde{\cal M}$ can be made larger as represented by the graded red regions, but only as long as ${\rm dim}\,\tilde{\cal M}$ is sufficiently smaller than ${\rm dim}\,{\cal M}$.
 This implies that a single $\tilde{\cal M}$ cannot cover a significant portion of ${\cal M}$.
 Note that the figure is only a schematic representation of the situation.}
\label{fig:M}
\end{figure}

Recall that the space of vacuum microstates $\tilde{\cal M}$ which these operators cover need only satisfy Eq.~(\ref{eq:tilde-M_1}).
In particular, it need not be the minimal choice given in Eq.~(\ref{eq:tilde-M_2}).
This implies that by increasing $S_{\rm eff}$, we can cover more and more microstates by a fixed set of infalling mode operators.
In fact, since the error $\epsilon$ of using these operators is given by Eq.~(\ref{eq:epsilon}), we can take any $S_{\rm eff}$ with
\begin{equation}
  S_{\rm eff} = c\, \{ S_{\rm bh}(M) + S_{\rm rad} \}
\qquad
  (0 < c < 1),
\label{eq:local-Seff}
\end{equation}
while keeping the error to be exponentially suppressed in the coarse-grained entropy of the system, unless $c$ is exponentially close to $1$.
In other words, within this $e^{S_{\rm eff}}$-dimensional vacuum microstate space $\tilde{\cal M}$, we can use the fixed operators ${\cal A}_\xi$ and ${\cal A}_\xi^\dagger$ throughout; i.e., these operators act state independently in this subspace of ${\cal M}$ (see Fig.~\ref{fig:M}).

It might seem that this allows us to construct ``fully global'' state independent operators which can be applied to most of the states in ${\cal M}$, since we can take $c$ in Eq.~(\ref{eq:local-Seff}) to be very close to $1$ for a macroscopic black hole.
This is, however, not the case.
The dimension of the space $\tilde{\cal M}_\perp$ of vacuum microstates that are orthogonal to the states in $\tilde{\cal M}$ is
\begin{equation}
  {\rm dim}\, \tilde{\cal M}_\perp = e^{S_{\rm tot}}-e^{S_{\rm eff}},
\end{equation}
which is much larger than ${\rm dim}\, \tilde{\cal M}$ even for $c$ close to $1$ (unless it is exponentially close).
In fact, for $c > 1/2$, there is a simple relation between the fraction of ${\cal M}$ which a fixed set of operators can cover and the size of error for using these operators:
\begin{equation}
  \frac{{\rm dim}\, \tilde{\cal M}}{{\rm dim}\, {\cal M}} \,\approx\, \epsilon.
\end{equation}
Therefore, if we want to keep the error small, $\epsilon \ll 1$, then the set of operators can be used only for a small fraction of states in ${\cal M}$.

It is instructive to see how a fixed set of infalling mode operators ${\cal A}_\xi$ and ${\cal A}_\xi^\dagger$ with $c < 1$ fails for a generic microstate, whose corresponding vacuum is not contained in $\tilde{\cal M}$.
Let $A'$ and $\bar{A}$ label orthonormal basis states of $\tilde{\cal M}$ and $\tilde{\cal M}_\perp$, respectively:
\begin{equation}
  A' = 1, \cdots, e^{S_{\rm eff}},
\qquad
  \bar{A} = 1, \cdots, e^{S_{\rm tot}}-e^{S_{\rm eff}}.
\label{eq:A-indices}
\end{equation}
We consider a generic mirror microstate having a specific semiclassical configuration $\{ \lambda_\alpha \}$:
\begin{equation}
  \ketc{\{ \lambda_\alpha \}} = \sum_{A'=1}^{e^{S_{\rm eff}}} a_{A'} \ketc{\{ \lambda_\alpha \}_{A'}} + \sum_{\bar{A}=1}^{e^{S_{\rm tot}}-e^{S_{\rm eff}}} \bar{a}_{\bar{A}} \ketc{\{ \lambda_\alpha \}_{\bar{A}}},
\label{eq:generic}
\end{equation}
where $\sum_{A'=1}^{e^{S_{\rm eff}}} |a_{A'}|^2 + \sum_{\bar{A}=1}^{e^{S_{\rm tot}}-e^{S_{\rm eff}}} |\bar{a}_{\bar{A}}|^2 = 1$.
The coefficients $a_{A'}$ and $\bar{a}_{\bar{A}}$ then have the statistical properties
\begin{equation}
  \vev{a_{A'}} = \vev{\bar{a}_{\bar{A}}} = 0,
\qquad
  \sqrt{\vev{|a_{A'}|^2}} = \sqrt{\vev{|\bar{a}_{\bar{A}}|^2}} = \frac{1}{e^{\frac{1}{2}S_{\rm tot}}}
\end{equation}
with uniformly distributed phases, where $\vev{\cdots}$ represents an average over microstate indices.

Let us now calculate the matrix elements of products of operators ${\cal A}_\xi$ and ${\cal A}_\xi^\dagger$, denoted by ${\cal O} = {\cal O}(\{ {\cal A}_\xi, {\cal A}_\xi^\dagger \})$, between generic states in Eq.~(\ref{eq:generic}).
They are given by
\begin{align}
  & \brac{\{ \kappa_\alpha \}} {\cal O} \ketc{\{ \lambda_\alpha \}} = \sum_{A',B'=1}^{e^{S_{\rm eff}}} a_{A'}^* a_{B'} \brac{\{ \kappa_\alpha \}_{A'}} {\cal O} \ketc{\{ \lambda_\alpha \}_{B'}} 
\nonumber\\*
  & \qquad + \left( \sum_{A'=1}^{e^{S_{\rm eff}}} \sum_{\bar{B}=1}^{e^{S_{\rm tot}}-e^{S_{\rm eff}}} a_{A'}^* \bar{a}_{\bar{B}} \brac{\{ \kappa_\alpha \}_{A'}} {\cal O} \ketc{\{ \lambda_\alpha \}_{\bar{B}}} + {\rm h.c.} \right) + \sum_{\bar{A},\bar{B}=1}^{e^{S_{\rm tot}}-e^{S_{\rm eff}}} \bar{a}_{\bar{A}}^* \bar{a}_{\bar{B}} \brac{\{ \kappa_\alpha \}_{\bar{A}}} {\cal O} \ketc{\{ \lambda_\alpha \}_{\bar{B}}}.
\end{align}
Using Eq.~(\ref{eq:correction}), we find that the first term on the right-hand side gives
\begin{equation}
  \sum_{A',B'=1}^{e^{S_{\rm eff}}} a_{A'}^* a_{B'} \brac{\{ \kappa_\alpha \}_{A'}} {\cal O} \ketc{\{ \lambda_\alpha \}_{B'}} = \sum_{A'=1}^{e^{S_{\rm eff}}} |a_{A'}|^2 {\cal M}^{\cal O}_{\kappa\lambda} \left[ 1 + O\left( {\rm max}\left\{\frac{e^{S_{\rm eff}}}{e^{S_{\rm bh}(M)+S_{\rm rad}}}, \frac{1}{e^{\frac{1}{2}\{S_{\rm bh}(M)+S_{\rm rad}\}}}\right\} \right) \right],
\label{eq:potential}
\end{equation}
where ${\cal M}^{\cal O}_{\kappa\lambda}$ represents the corresponding matrix elements in field theory.
This, therefore, gives the correct matrix elements up to the universal overall factor of $\sum_{A'=1}^{e^{S_{\rm eff}}} |a_{A'}|^2 = O(e^{S_{\rm eff}}/e^{S_{\rm tot}})$, which could be absorbed into the normalization of states and hence does not affect physics.

What about the second and third terms?
To figure this out, we can use relations analogous to Eq.~(\ref{eq:correction}) in which one or both of the ``outermost'' indices are replaced with barred ones:
\begin{align}
  & \sum_{A_1'=1}^{e^{S_{\rm eff}}} \sum_{A_2'=1}^{e^{S_{\rm eff}}} \cdots \sum_{A_{p-1}'=1}^{e^{S_{\rm eff}}} \eta_{\lambda_1}^{E'A_1'} \eta_{\lambda_2}^{A_1'A_2'} \cdots \eta_{\lambda_p}^{A_{p-1}'\bar{F}} 
  = \varepsilon_{\lambda_p}^{E'\bar{F}},
\\*
  & \sum_{A_1'=1}^{e^{S_{\rm eff}}} \sum_{A_2'=1}^{e^{S_{\rm eff}}} \cdots \sum_{A_{p-1}'=1}^{e^{S_{\rm eff}}} \eta_{\lambda_1}^{\bar{E}A_1'} \eta_{\lambda_2}^{A_1'A_2'} \cdots \eta_{\lambda_p}^{A_{p-1}'\bar{F}} 
  = \sum_{A_1'=1}^{e^{S_{\rm eff}}} \varepsilon_{\lambda_1}^{\bar{E}A_1'} \varepsilon_{\lambda_p}^{A_1'\bar{F}}.
\label{eq:correction-2}
\end{align}
We find that the second term is suppressed by a factor of $O(e^{\frac{1}{2}S_{\rm eff}}/e^{S_{\rm tot}})$ relative to the first term, and hence negligible.
The third term, however, contains a contribution
\begin{align}
  & \sum_{\bar{A},\bar{B}=1}^{e^{S_{\rm tot}}-e^{S_{\rm eff}}} \bar{a}_{\bar{A}}^* \bar{a}_{\bar{B}}\, {\cal M}^{\cal O}_{\kappa\lambda} z^2 e^{\frac{E_\kappa+E_\lambda}{2T_{\rm H}}} \sum_{D'=1}^{e^{S_{\rm eff}}} \left( \sum_{i_\kappa = 1}^{e^{S_{\rm bh}(M-E_\kappa)}} \sum_{a = 1}^{e^{S_{\rm rad}}} c^{\bar{A}*}_{\kappa i_\kappa a} c^{D'}_{\kappa i_\kappa a} \right) \left( \sum_{j_\lambda = 1}^{e^{S_{\rm bh}(M-E_\lambda)}} \sum_{b = 1}^{e^{S_{\rm rad}}} c^{D' *}_{\lambda j_\lambda b} c^{\bar{B}}_{\lambda j_\lambda b} \right) \delta_{\bar{A}\bar{B}}
\nonumber\\*
  &\qquad\qquad = \sum_{\bar{A}=1}^{e^{S_{\rm tot}}-e^{S_{\rm eff}}} |\bar{a}_{\bar{A}}|^2 {\cal M}^{\cal O}_{\kappa\lambda}\, O\left( \frac{e^{S_{\rm eff}}}{e^{S_{\rm bh}(M)+S_{\rm rad}}} \right) \delta_{\kappa\lambda} + \cdots,
\label{eq:3rd-term}
\end{align}
in which the factor multiplied to ${\cal M}^{\cal O}_{\kappa\lambda}$ depends on the external states.
This contribution (and only this contribution) is comparable to the first term, jeopardizing the potential success of Eq.~(\ref{eq:potential}).
This lack of fully global operators is the state dependence discussed in Refs.~\cite{Papadodimas:2013jku,Papadodimas:2015jra}.

An analysis similar to the one above shows that the state independent operators in $\tilde{\cal M}$ works correctly for a typical state built on vacuum microstate space $\tilde{\cal M}' \supset \tilde{\cal M}$ as long as
\begin{equation}
  {\rm dim}\,\tilde{\cal M}' - {\rm dim}\,\tilde{\cal M} \,<\, {\rm dim}\,\tilde{\cal M},
\end{equation}
with the error of order
\begin{equation}
  \epsilon' = \frac{{\rm dim}\, \tilde{\cal M}' - {\rm dim}\, \tilde{\cal M}}{{\rm dim}\, \tilde{\cal M}}.
\end{equation}
This implies that the condition of Eq.~(\ref{eq:tilde-M_1}) is actually a little weaker.
For a typical state in the Schmidt basis, $\tilde{\cal M}$ can be smaller than $\tilde{V}[\ket{\Psi(t_*)}]$ as long as
\begin{equation}
  {\rm dim}\,\tilde{V}\bigl[\ket{\Psi(t_*)}\bigr] - {\rm dim}\,\tilde{\cal M} \,<\, {\rm dim}\,\tilde{\cal M}.
\end{equation}

The space $\tilde{\cal M}$ is analogous to a code subspace of the quantum error correction interpretation of holography described in Ref.~\cite{Almheiri:2014lwa}.
One difference is that there is no preferred choice of the code subspace determined by semiclassical geometries.
At the semiclassical level, all the states in ${\cal M}$ look like having the same spacetime with the same black hole.
We can therefore naturally take any $e^{S_{\rm eff}}$ independent vacuum microstates satisfying Eq.~(\ref{eq:local-Seff}) to form $\tilde{\cal M}$.%
\footnote{Because of this freedom for choosing a set of infalling operators, one might think that we can cover the entire ${\cal M}$ space by exponentially large number, $O(e^{S_{\rm tot}-S_{\rm eff}})$, of fixed such sets.
 This is, however, not the case; using the argument through Eqs.~(\ref{eq:A-indices}~--~\ref{eq:3rd-term}), it is easy to see that even a typical state in ${\cal M}$ is not covered by any of these sets.
 To cover all states in ${\cal M}$ by fixed sets of operators, we need double exponentially large number, $O(e^{e^{S_{\rm tot}-S_{\rm eff}}})$, of sets.
 This is related to the well-known fact that in a Hilbert space of dimension $e^S \gg 1$, there are $O(e^{e^S})$ approximately orthogonal states with exponentially small overlaps of $O(e^{-\frac{1}{2}S})$.}
A similar aspect of quantum error correction in spacetime with a black hole has been analyzed in Ref.~\cite{Hayden:2018khn}.

\subsection{A young black hole and the Minkowski limit}
\label{subsec:comments}

So far, we have considered the effective theory of the interior erected using mirror operators constructed out of the soft and far modes.
However, for a young black hole, i.e.\ a black hole that is not yet maximally entangled with the rest of the system, these operators can be constructed only out of the soft modes, using the so-called Petz map~\cite{Penington:2019kki,Nomura:2019dlz}:%
\footnote{The Petz map was used in Ref.~\cite{Penington:2019kki} to construct interior operators acting only on radiation.
This was possible because the analysis did not consider the energy constraint imposed on the black hole system, i.e.\ the hard and soft modes; with the energy constraint, such a construction is not possible~\cite{Nomura:2019dlz}.
 Note also that the expression for the operators given in Ref.~\cite{Nomura:2019dlz} adopted the so-called the Petz-light map~\cite{Penington:2019kki}; for the purpose of this paper, however, we need to use the full Petz map, given explicitly below.}
\begin{equation}
  \tilde{\cal O}^A[{\cal O}] = \sum_\kappa \sum_\lambda {\cal O}_{\kappa\lambda}\, \alpha^A_\kappa \alpha^{A*}_\lambda \sum_{i_\kappa,i'_\kappa=1}^{e^{S_{\rm bh}(M-E_\kappa)}} \sum_{j_\lambda,j'_\lambda=1}^{e^{S_{\rm bh}(M-E_\lambda)}} \sum_{a=1}^{e^{S_{\rm rad}}} X^{(\kappa,A)}_{i'_\kappa i_\kappa} c^A_{\kappa i_\kappa a} c^{A*}_{\lambda j_\lambda a} X^{(\lambda,A)}_{j_\lambda j'_\lambda} \ket{\psi^{(\kappa)}_{i'_\kappa}} \bra{\psi^{(\lambda)}_{j'_\lambda}},
\end{equation}
where $A = 1,\cdots,e^{S_{\rm tot}}$ labels microstates, $\alpha^A_\kappa$ is given by Eq.~(\ref{eq:alpha_nA}), and
\begin{equation}
  X^{(\kappa,A)}_{i_\kappa j_\kappa} = e^{-\frac{1}{2}S_{\rm rad}}\! \sum_{a=1}^{e^{S_{\rm rad}}} \frac{c^A_{\kappa i_\kappa a} c^{A*}_{\kappa j_\kappa a}}{\left| \alpha^A_\kappa \sum_{k_\kappa=1}^{e^{S_{\rm bh}(M-E_\kappa)}}\! c^{A*}_{\kappa k_\kappa a} c^A_{\kappa k_\kappa a} \right|^2}.
\label{eq:X}
\end{equation}
The matrix elements ${\cal O}_{\kappa\lambda}$ are those of operator ${\cal O}$ in field theory; for mirror annihilation and creation operators
\begin{alignat}{5}
  {\cal O}_{\kappa\lambda} &= \sqrt{\lambda_\gamma}\, \delta_{\kappa \lambda_\smg}
\;\; && \mbox{for } {\cal O} = \tilde{b}_\gamma 
  && \quad\Rightarrow\quad \tilde{\cal O}^A[{\cal O}] = \tilde{b}_\gamma^A,
\\*
  {\cal O}_{\kappa\lambda} &= \sqrt{\lambda_\gamma+1}\, \delta_{\kappa \lambda_\spg}
\;\; && \mbox{for } {\cal O} = \tilde{b}_\gamma^\dagger 
  && \quad\Rightarrow\quad \tilde{\cal O}^A[{\cal O}] = \tilde{b}_\gamma^{A\dagger}.
\end{alignat}

The quantity $X^{(\kappa,A)}_{i_\kappa j_\kappa}$ in Eq.~(\ref{eq:X}) satisfies
\begin{align}
  & \sum_{i_\kappa,j_\kappa=1}^{e^{S_{\rm bh}(M-E_\kappa)}}\!\!\! \alpha^{A*}_\kappa c^{A*}_{\kappa i_\kappa a} [X^{(\kappa,A)} X^{(\kappa,B)}]_{i_\kappa j_\kappa} \alpha^B_\kappa c^B_{\kappa j_\kappa b}
\nonumber\\*
  &\qquad\qquad = \delta_{AB} \delta_{ab} \left\{ 1 + O\biggl(\frac{1}{e^{\frac{1}{2}S{\rm rad}}}\biggr) + O\biggl(\frac{e^{S_{\rm rad}}}{e^{S_{\rm bh}(M)}}\biggr) \right\} + O\biggl(\frac{1}{e^{\frac{1}{2}S_{\rm bh}(M)}}\biggr),
\label{eq:relation-1}
\end{align}
where we have kept only the terms that can dominate for $S_{\rm rad} < S_{\rm bh}(M)$.
It also satisfies
\begin{align}
  & e^{\frac{1}{2}S_{\rm rad}} \sum_{i_\kappa,j_\kappa=1}^{e^{S_{\rm bh}(M-E_\kappa)}} \alpha^{D*}_\kappa c^{D*}_{\kappa i_\kappa a} X^{(\kappa,A)}_{i_\kappa j_\kappa} \alpha^E_\kappa c^E_{\kappa j_\kappa b}
\nonumber\\*
  &\qquad\qquad = \begin{cases}
    \delta_{ab} \left\{ 1 + O\Bigl(\frac{e^{S_{\rm rad}}}{e^{S_{\rm bh}(M)}}\Bigr) \right\} + (1-\delta_{ab}) O\Bigl(\frac{1}{e^{\frac{1}{2}S_{\rm bh}(M)}}\Bigr) & \mbox{for } D = A = E
\\
  O\Bigl(\frac{1}{e^{\frac{1}{2}S_{\rm bh}(M)}}\Bigr) & \mbox{for } D = A \neq E \mbox{ or } D \neq A = E
\\
  \delta_{ab} O\Bigl(\frac{e^{S_{\rm rad}}}{e^{S_{\rm bh}(M)}}\Bigr) + O\Bigl(\frac{e^{\frac{1}{2}S_{\rm rad}}}{e^{S_{\rm bh}(M)}}\Bigr) & \mbox{for } D = E \neq A
\\
  O\Bigl(\frac{e^{\frac{1}{2}S_{\rm rad}}}{e^{S_{\rm bh}(M)}}\Bigr) & \mbox{for } D, A, E \mbox{ all different}.
  \end{cases}
\label{eq:relation-2}
\end{align}
We have again kept only terms that can dominate for $S_{\rm rad} < S_{\rm bh}(M)$.

The relation in Eq.~(\ref{eq:relation-1}) implies that the product of microscopic operators corresponding to field theory operators ${\cal O}_1$ and ${\cal O}_2$ is the microscopic operator corresponding to the field theory operator ${\cal O}_1 {\cal O}_2$ up to exponentially suppressed corrections:
\begin{equation}
  \tilde{\cal O}^A[{\cal O}_1]\, \tilde{\cal O}^B[{\cal O}_2] = \delta_{AB}\, \tilde{\cal O}^A[{\cal O}_1 {\cal O}_2] \left\{ 1 + O\biggl(\frac{1}{e^{\frac{1}{2}S{\rm rad}}}\biggr) + O\biggl(\frac{e^{S_{\rm rad}}}{e^{S_{\rm bh}(M)}}\biggr) \right\} + O\biggl(\frac{1}{e^{\frac{1}{2}S_{\rm bh}(M)}}\biggr).
\end{equation}
Note that this desired property persists only for a young black hole, $S_{\rm rad} < S_{\rm bh}(M)$, since otherwise the corrections are not exponentially suppressed~\cite{Nomura:2019dlz}.
The relation in Eq.~(\ref{eq:relation-2}) implies that the matrix element of a microscopic operator $\tilde{\cal O}^A[{\cal O}]$ between mirror microstates with the same microstate index is the same as that in field theory up to exponentially suppressed corrections:
\begin{equation}
  \brac{\{ \kappa_\alpha \}_D} \tilde{\cal O}^A[{\cal O}] \ketc{\{ \lambda_\alpha \}_E}
  = {\cal O}_{\kappa\lambda} \times \begin{cases}
    1 + O\Bigl(\frac{e^{S_{\rm rad}}}{e^{S_{\rm bh}(M)}}\Bigr) + O\Bigl(\frac{1}{e^{\frac{1}{2}S_{\rm bh}(M)}}\Bigr) & \mbox{for } D = A = E
\\
  O\Bigl(\frac{1}{e^{\frac{1}{2}S_{\rm bh}(M)}}\Bigr) & \mbox{for } D = A \neq E \mbox{ or } D \neq A = E
\\
  \delta_{\kappa\lambda} O\Bigl(\frac{e^{S_{\rm rad}}}{e^{S_{\rm bh}(M)}}\Bigr) + O\Bigl(\frac{1}{e^{S_{\rm bh}(M)}}\Bigr) & \mbox{for } D = E \neq A
\\
  O\Bigl(\frac{1}{e^{S_{\rm bh}(M)}}\Bigr) & \mbox{for } D, A, E \mbox{ all different}.
  \end{cases}
\label{eq:mat-el}
\end{equation}

The expression in Eq.~(\ref{eq:mat-el}) allows us to analyze global promotion of operators $\tilde{\cal O}^A[{\cal O}]$ analogously to the case of operators comprising both soft and far modes.
Specifically, we can consider globally promoted operators
\begin{equation}
  \tilde{\cal O} = \sum_{A'=1}^{e^{S_{\rm eff}}} \tilde{\cal O}^{A'}[{\cal O}]
\label{eq:global-soft}
\end{equation}
and calculate the matrix elements between states built on space $\tilde{\cal M}$:
\begin{equation}
  \brac{\{ \kappa_\alpha \}_{A'}} \tilde{\cal O} \ketc{\{ \lambda_\alpha \}_{B'}}
  = {\cal O}_{\kappa\lambda} \times \begin{cases}
    1 + O\Bigl(\frac{e^{S_{\rm rad}} e^{S_{\rm eff}}}{e^{S_{\rm bh}(M)}}\Bigr) + O\Bigl(\frac{1}{e^{\frac{1}{2}S_{\rm bh}(M)}}\Bigr) & \mbox{for } A' = B'
\\
  O\Bigl(\frac{1}{e^{\frac{1}{2}S_{\rm bh}(M)}}\Bigr) + O\Bigl(\frac{e^{\frac{1}{2}S_{\rm eff}}}{e^{S_{\rm bh}(M)}}\Bigr) & \mbox{for } A' \neq B'.
  \end{cases}
\end{equation}
This implies that the error of globally promoted operators is of order
\begin{equation}
  \epsilon_\psi = {\rm max}\left\{ \frac{e^{S_{\rm rad}+S_{\rm eff}}}{e^{S_{\rm bh}(M)}}, \frac{1}{e^{\frac{1}{2}S_{\rm bh}(M)}} \right\},
\label{eq:eps-psi}
\end{equation}
so that the globally promoted operators work as long as
\begin{equation}
  S_{\rm eff} < S_{\rm bh}(M) - S_{\rm rad}
\end{equation}
(unless $S_{\rm eff}$ is exponentially close to $S_{\rm bh}(M) - S_{\rm rad}$).
This agrees with the result of the general analysis in Ref.~\cite{Hayden:2018khn}.

\subsubsection*{The Minkowski limit}

We finally make a brief comment on the Minkowski limit.
This limit is obtained by taking $M \rightarrow \infty$ with the other physical scales fixed and focusing on the near horizon region.
As discussed in Ref.~\cite{Nomura:2019qps}, the limit is smooth.
Taking the exterior view, we recover Rindler space with the correct implication that we cannot retrieve information from Unruh radiation in any finite time, and making the horizon recede returns an object behind the horizon to the front without it being scrambled.
What is the implication of this limit for the operators we have discussed?

Assuming that semiclassical matter we want to describe did not exist for an infinitely long time, we expect that $S_{\rm eff}$ is finite.
With this assumption, we find that Eq.~(\ref{eq:epsilon}), or Eq.~(\ref{eq:eps-psi}) with $S_{\rm rad} < S_{\rm bh}(M)$ at the leading order, gives $\epsilon \rightarrow 0$ as $S_{\rm bh}(M) \rightarrow \infty$ under $M \rightarrow \infty$.
This implies that (globally promoted) Minkowski operators describing the physics of semiclassical matter need not have an error analogous to Eq.~(\ref{eq:epsilon}).
In fact, the error for a black hole comes from the fact that the system, of which the semiclassical matter is a part, has only a finite number of degrees of freedom.%
\footnote{In this respect, the zone region of a black hole is viewed as a finite system (coupled weakly to the external system) because of the gravitational confining potential; see Eqs.~(\ref{eq:non-ortho-1},~\ref{eq:non-ortho-2}).}

\section{Conclusion}
\label{sec:concl}

From the point of view of quantum gravity, what is special about a system with a horizon is that nonperturbative gauge redundancies of a gravitational theory become so prominent that there are widely different descriptions of the same physics.
In this paper, we have studied the unitary gauge construction---which keeps unitarity manifest a la holography---of a system with a black hole.

We have presented an explicit procedure to construct quantum operators describing the interior of the black hole, which does not rely on details of microscopic physics.
A key role is played by semiclassical (hard) modes in the black hole zone, determining the relevant degrees of freedom in the soft and far modes, whose collective excitations play the role of those in the second exterior of the effective two-sided geometry.
We have analyzed the structure of infalling operators in the space of black hole vacuum microstates, including their state (in)dependence and a connection to the quantum error correction interpretation.

Since there are other physical systems possessing horizons, we expect that the picture developed here is pertinent in these cases as well, most notably in inflationary cosmology as explored in Refs.~\cite{Nomura:2019qps,Nomura:2011dt}.
It is our hope that the analysis in this paper provides a further clue to uncover how quantum gravity works for such systems.

\vspace{0.3cm}
\subsubsection*{Note added:}
\vspace{-0.2cm}

While completing this paper we received Ref.~\cite{Akal:2020ujg}, which seems to discuss a picture similar to that developed in Refs.~\cite{Langhoff:2020jqa,Nomura:2018kia,Nomura:2019qps,Nomura:2019dlz} and in this paper.

\section*{Acknowledgments}

This work was supported in part by the Department of Energy, Office of Science, Office of High Energy Physics under contract DE-AC02-05CH11231 and award DE-SC0019380.

\appendix

\section{Consistency with the Semiclassical Expectation}
\label{app:semi}

In this appendix, we show that the scrambling dynamics of a black hole in the exterior view is consistent with what we expect in the interior in the semiclassical picture.
Specifically, we consider the situation in which a falling object, as viewed from the exterior, reaches the stretched horizon at $t = t_*$, and ask what we expect to see in the effective theory erected at $t = t_* + \varDelta t$ with $\varDelta t \geq 0$.

For concreteness, let us consider a black hole in $(d+1)$-dimensional asymptotically flat spacetime (or a small black hole in $(d+1)$-dimensional asymptotically AdS spacetime) with $d \geq 3$:
\begin{equation}
  ds^2 = -\left( 1 - \frac{r_+^{d-2}}{r^{d-2}} \right) dt^2 + \frac{1}{1 - \frac{r_+^{d-2}}{r^{d-2}}}\, dr^2 + r^2 d\Omega_{d-1}^2,
\end{equation}
where $r_+$ is the horizon radius.
The mass, temperature, and entropy of the black hole are given by
\begin{equation}
  M = (d-1) \frac{r_+^{d-2}}{16 \pi l_{\rm P}^{d-1}}\, {\rm vol}(\Omega_{d-1}),
\qquad
  T_{\rm H} = \frac{d-2}{4\pi r_+},
\qquad
  S_{\rm bh} = \frac{r_+^{d-1}}{4 l_{\rm P}^{d-1}}\, {\rm vol}(\Omega_{d-1}),
\end{equation}
where $l_{\rm P}$ is the $(d+1)$-dimensional Planck length, and ${\rm vol}(\Omega_{d-1}) = 2\pi^{d/2}/\Gamma(d/2)$ is the volume of the $(d-1)$-dimensional unit sphere.
The stretched horizon and the edge of the zone are located at~\cite{Langhoff:2020jqa}
\begin{equation}
  r_{\rm s} - r_+ \approx \frac{d-2}{4} \frac{l_{\rm s}^2}{r_+},
\qquad
  r_{\rm z} \approx \left(\frac{d}{2}\right)^{\frac{1}{d-2}} r_+,
\end{equation}
respectively, where $l_{\rm s}$ is the string length.

\begin{figure}[t]
\centering
  \subfigure[]{\includegraphics[trim = 1.5cm 0 0 0,clip,height=5.8cm]{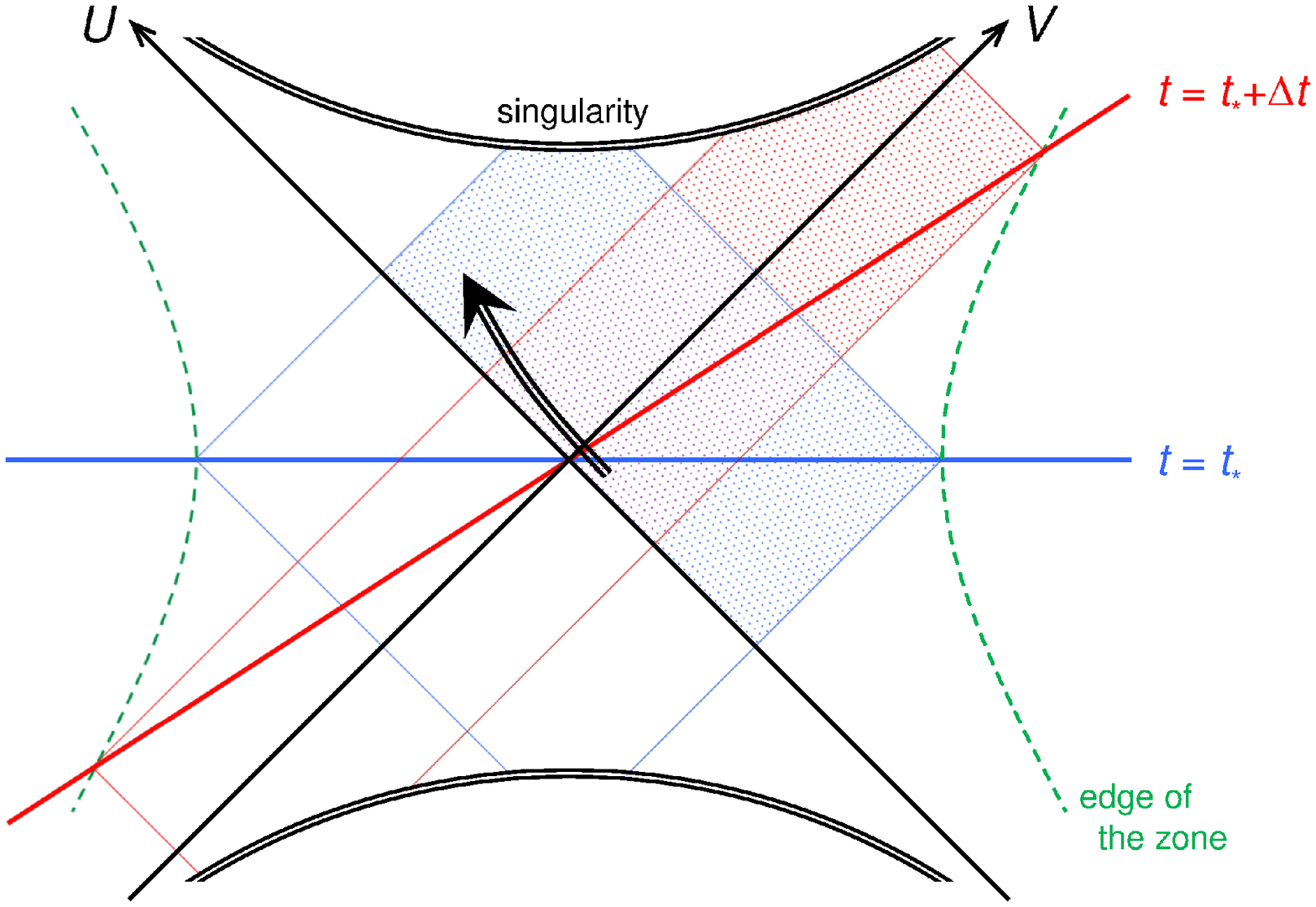}}
\hspace{0.3cm}
  \subfigure[]{\includegraphics[trim = 1.5cm 0 0 0,clip,height=5.8cm]{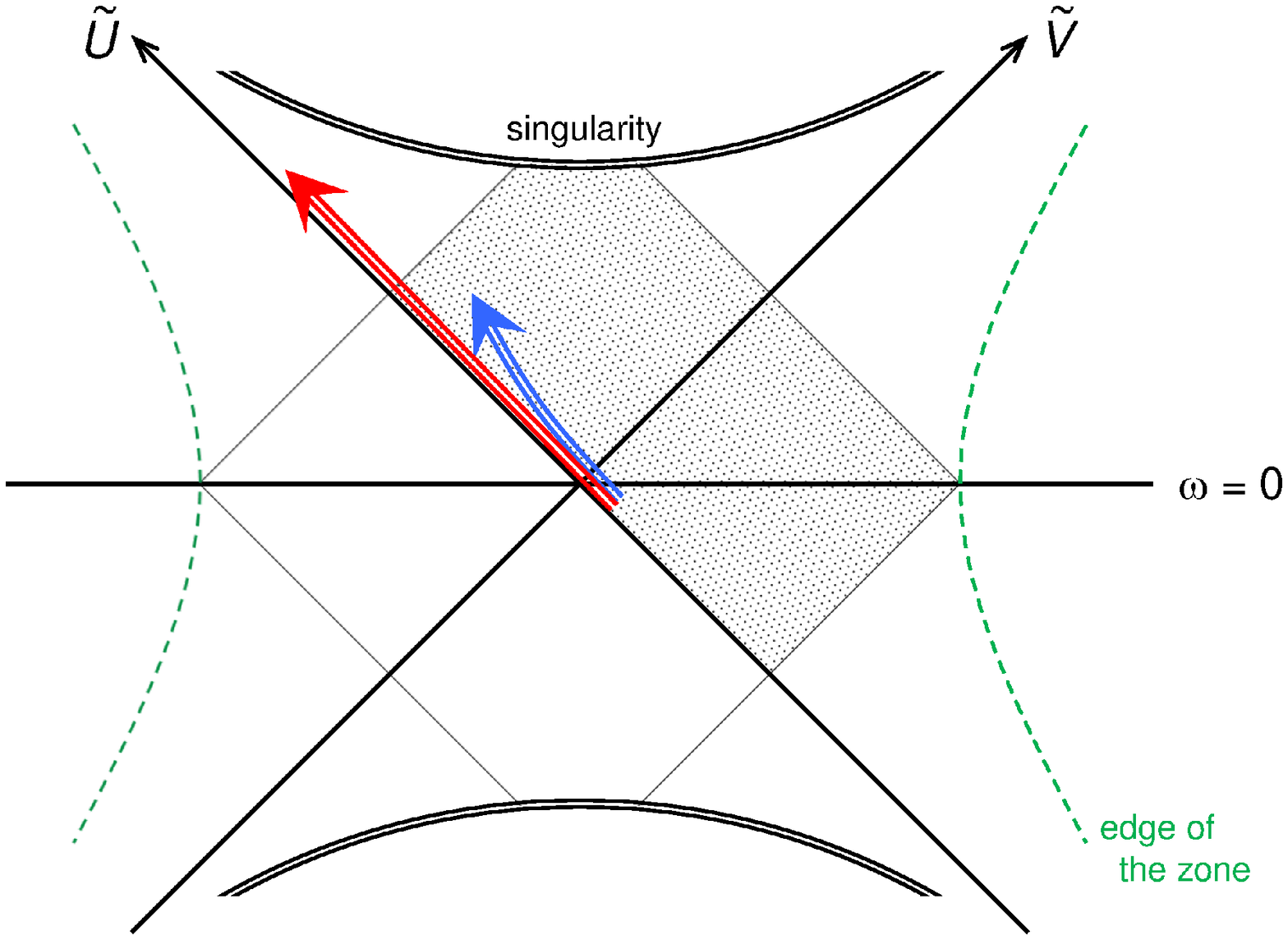}}
\caption{(a) To describe the fate of a falling object that reached the stretched horizon at $t = t_*$, we may erect an effective theory of the interior at $t = t_*$ (blue) or some time later $t = t_* + \varDelta t$ with $\varDelta t > 0$ (red).
 (b) The same is depicted in the coordinates adapted to each effective theory, $\tilde{U}$ and $\tilde{V}$.
 The blue and red arrows represent the trajectory of the object in the effective theories erected at $t = t_*$ (blue) and $t = t_* + \varDelta t$ (red), respectively.}
\label{fig:eff-ths}
\end{figure}

To describe the fate of the object reaching the stretched horizon at $t = t_*$, we can erect an effective theory of the interior at that time, $t = t_*$.
This theory has the geometry of a two-sided black hole
\begin{equation}
  ds^2 = -dU dV + r^2 d\Omega_{d-1}^2,
\end{equation}
where $U$ and $V$ are the Kruskal-Szekeres coordinates given by
\begin{eqnarray}
  \begin{cases}
    U = -R\, e^{-\omega} \\
    V = R\, e^{\omega}
  \end{cases}
\label{eq:KS-1}
\end{eqnarray}
in the first exterior (i.e.\ Region~I:\ $U < 0$ and $V > 0$) and by
\begin{eqnarray}
  \begin{cases}
    U = R\, e^{-\omega} \\
    V = R\, e^{\omega},
  \end{cases}
\qquad
  \begin{cases}
    U = R\, e^{-\omega} \\
    V = -R\, e^{\omega},
  \end{cases}
\qquad
  \begin{cases}
    U = -R\, e^{-\omega} \\
    V = -R\, e^{\omega}
  \end{cases}
\label{eq:KS-2}
\end{eqnarray}
in the other regions:\ Region~II ($U, V > 0$), Region~III ($U > 0, V < 0$), and Region~IV ($U, V < 0$), respectively.
Here, $R$ and $\omega$ are given by
\begin{eqnarray}
  \begin{cases}
    R = \frac{2r_+}{\sqrt{d-2}} \sqrt{\frac{|r-r_+|}{r_+}} \\
    \omega = \frac{d-2}{2r_+}(t-t_*)
  \end{cases}
\label{eq:R-omega}
\end{eqnarray}
in the near horizon approximation.%
\footnote{To obtain the explicit expression in Eq.~(\ref{eq:R-omega}), we need to use the near horizon approximation $|r - r_+| \ll r_+$.
 The results described here, however, are valid at an order of magnitude level even for $|r - r_+| \approx O(r_+)$.}
This is illustrated by the blue diamond region in Fig.~\ref{fig:eff-ths}(a), where the physically relevant region (the first exterior and the interior) is shaded.

The effective theory, however, need not be erected right at $t = t_*$ but can be erected at some time later $t = t_* + \varDelta t$ ($\varDelta t > 0$) as illustrated by red color in Fig.~\ref{fig:eff-ths}(a).
If we erect an effective theory soon enough, then we still find the object in the interior region in a way consistent with the semiclassical expectation.
As discussed in Section~\ref{subsec:match}, this is possible because the black hole state is not yet equilibrated, so that collective excitations of the soft modes represent the existence of the object in the interior.
On the other hand, if we erect an effective theory more than the scrambling time after $t = t_*$, i.e.\ $\varDelta t > t_{\rm scr}$, then the effective theory will find that the interior is in the semiclassical vacuum, since the black hole state is equilibrated by then.
Is this consistent with what we expect from the semiclassical picture?

Let us consider the Kruskal-Szekeres coordinates $\tilde{U}$ and $\tilde{V}$ of the effective theory erected at $t = t_* + \varDelta t$.
These are related to $U$ and $V$ in Eqs.~(\ref{eq:KS-1}~--~\ref{eq:R-omega}) by
\begin{equation}
\begin{cases}
  \tilde{U} &= U e^{\frac{d-2}{2r_+}\varDelta t} \\
  \tilde{V} &= V e^{-\frac{d-2}{2r_+}\varDelta t}
\end{cases}
\label{eq:KS-rel}
\end{equation}
(so that $\tilde{U} = U$ and $\tilde{V} = V$ for $\varDelta t = 0$).
For $\varDelta t > 0$, the physical spacetime region that can be described by this theory is given as the red shaded region in Fig.~\ref{fig:eff-ths}(a), so it must satisfy
\begin{equation}
  \tilde{U} \lesssim r_{\rm z} = O(r_+)
\quad\Rightarrow\quad
  U \lesssim r_+ e^{-\frac{d-2}{2r_+}\varDelta t}.
\label{eq:U-bound}
\end{equation}
This implies that the object that fell into the horizon at $t = t_*$ can spend the proper time of only
\begin{equation}
  \varDelta \tau \lesssim r_+ e^{-\frac{d-2}{2r_+}\varDelta t}
\end{equation}
in this region.
Since holography limits that the maximum amount of information an object can handle is of $O(1)$ per Planck time, this implies that the object cannot cause any physical effect in an effective theory erected after $t = t_* + \varDelta t_{\rm max}$, with
\begin{equation}
  \varDelta t_{\rm max} = \frac{2r_+}{d-2} \ln\frac{r_+}{l_{\rm P}}.
\label{eq:Dt_max}
\end{equation}
Here, the fractional correction to this equation is only of order $1/\ln(r_+/l_{\rm P})$, which is also the case in all similar expressions below.

We now compare this time with the scrambling time.
The scrambling time of a $(d+1)$-dimensional flat space black hole was calculated in Ref.~\cite{Penington:2019npb} in ingoing Eddington-Finkelstein time as
\begin{equation}
  v_{\rm scr} = \frac{2(d-1)}{d-2} r_+ \ln\frac{r_+}{l_{\rm P}}.
\label{eq:v_scr}
\end{equation}
This is related to the scrambling time relevant here in boundary time by
\begin{equation}
  t_{\rm scr} = v_{\rm scr} - t_{\rm sig} = 2r_+ \ln\frac{r_+}{l_{\rm P}},
\end{equation}
where $t_{\rm sig} = (2r_+/(d-2))\ln(r_+/l_{\rm P})$ is the signal propagation time~\cite{Langhoff:2020jqa}, and we have taken $l_{\rm s} \approx l_{\rm P}$.
(This is obtained by setting $r_{\rm e} \sim r_+$ in the expressions in Ref.~\cite{Langhoff:2020jqa}.)
We thus find that
\begin{equation}
  \frac{\varDelta t_{\rm max}}{t_{\rm scr}} = \frac{1}{d-2},
\label{eq:time-ratio}
\end{equation}
which is $1$ or smaller for $d \geq 3$.
In other words, if we erect the effective theory at the scrambling time after $t_*$, i.e.\ $\varDelta t = t_{\rm scr}$, then the maximal proper time which the object falling at $t = t_*$ can spend in the region described by the effective theory is
\begin{equation}
  \varDelta \tau = O(r_+) \left(\frac{l_{\rm P}}{r_+}\right)^{d-2},
\label{eq:del-tau}
\end{equation}
which is of $O(l_{\rm P})$ or smaller for $d \geq 3$.
Therefore, the object actually need not be included in effective theories erected after $t = t_* + t_{\rm scr}$, and the black hole interior in these theories should indeed be the semiclassical vacuum.
This, together with the fact that equilibration of black hole states takes time, comprises how the frozen vacuum problem of Ref.~\cite{Bousso:2013ifa} is addressed in the present framework.%
\footnote{The frozen vacuum is an example of the problems that are addressed trivially in the approach based on global spacetime but require nontrivial elaboration in the unitary gauge construction.
 An example of the opposite class of problems, which can be understood more easily in the unitary gauge construction, is the origin of the ensemble nature of Euclidean gravitational path integral found in the global spacetime picture~\cite{Penington:2019kki}.
 In the unitary gauge construction, this is understood simply as a statistics of the soft modes arising from the fact that these modes cannot be resolved at a timescale relevant for describing the black hole~\cite{Langhoff:2020jqa}.
 (If the gravitational---bulk---theory is two dimensional, then there is no ensemble of soft modes, since the horizon is spatially a point.
 In this case, however, the gravitational description emerges from an ensemble of unitary---boundary---theories~\cite{Saad:2019lba,Stanford:2019vob}, making a black hole state in the bulk correspond to an ensemble of microstates in these theories.
 A similar comment is expected to apply in three-dimensional pure gravity, which does not have propagating gravitons and hence soft modes.)
 More detailed discussion about the ``duality'' of the two constructions can be found in Ref.~\cite{Nomura:2020ewg}.}

There are a few comments.
First, the argument given here is essentially that in Ref.~\cite{Hayden:2007cs}, except that the relevant entity for the later time is the spacetime region of the effective theory, and not a physical observer per~se.
However, the analysis of Ref.~\cite{Hayden:2007cs} does not fix the coefficient of Eq.~(\ref{eq:v_scr}), which is crucial for the final conclusion because it determines the power of $l_{\rm P}/r_+$ in Eq.~(\ref{eq:del-tau}).
It is reassuring that the explicit result in Eq.~(\ref{eq:v_scr}) satisfies the necessary consistency condition, although the origin of the extra leeway---the factor of $d-2$ in Eqs.~(\ref{eq:time-ratio},~\ref{eq:del-tau})---is not clear.

Second, at the classical level, the trajectory of the object reaching the stretched horizon at $t_*$ is contained not only in the region described by the effective theory erected at $t = t_*$ but also in that erected at $t = t_* + \varDelta t$ for any $\varDelta t > 0$.
(In Fig.~\ref{fig:eff-ths}(b), we depict the trajectory of the object in two effective theories using the coordinates $\tilde{U}$ and $\tilde{V}$ adapted to each theory:\ blue for the theory erected at $t = t_*$ and red for that erected at $t = t_* + \varDelta t$.)
Nevertheless, at the quantum level, the object does not appear in effective theories erected at times after $t = t_* + t_{\rm scr}$.
This is the UV cutoff phenomenon discussed in Ref.~\cite{Nomura:2018kia} and is another manifestation of UV-IR relation in quantum gravity:\ the impossibility of communication between theories erected long time (IR) apart is ensured by the lack of short distance (UV) physics below the Planck length.

Finally, the analysis performed here can be extended straightforwardly to a large AdS black hole:
\begin{equation}
  ds^2 = -\left( \frac{r^2}{l^2} - \frac{r_+^d}{l^2 r^{d-2}} \right) dt^2 + \frac{1}{\frac{r^2}{l^2} - \frac{r_+^d}{l^2 r^{d-2}}}\, dr^2 + r^2 d\Omega_{d-1}^2,
\end{equation}
where $l$ is the AdS radius.
The properties of this black hole are given by%
\footnote{For the present purpose, we must take $r_{\rm z} = O(r_+)$, which is the largest length scale in the system.}
\begin{equation}
  M = (d-1) \frac{r_+^d}{16 \pi l_{\rm P}^{d-1} l^2}\, {\rm vol}(\Omega_{d-1}),
\qquad
  T_{\rm H} = \frac{d r_+}{4\pi l^2},
\qquad
  S_{\rm bh} = \frac{r_+^{d-1}}{4 l_{\rm P}^{d-1}}\, {\rm vol}(\Omega_{d-1}),
\end{equation}
\begin{equation}
  r_{\rm s} - r_+ \approx \frac{d}{4} \frac{r_+ l_{\rm s}^2}{l^2},
\qquad
  r_{\rm z} = O(r_+).
\end{equation}
The Kruskal-Szekeres coordinates $U$ and $V$ are now given (in the near horizon approximation) by Eqs.~(\ref{eq:KS-1},~\ref{eq:KS-2}) with
\begin{eqnarray}
  \begin{cases}
    R = \frac{2l}{\sqrt{d}} \sqrt{\frac{|r-r_+|}{r_+}} \\
    \omega = \frac{d r_+}{2 l^2}(t-t_*).
  \end{cases}
\label{eq:R-omega-AdS}
\end{eqnarray}
The appropriate modifications of Eq.~(\ref{eq:KS-rel}) and the first expression of Eq.~(\ref{eq:U-bound}) are then given by the replacements
\begin{equation}
  \frac{d-2}{r_+} \rightarrow \frac{d r_+}{l^2}
\end{equation}
and
\begin{equation}
  \tilde{U} \lesssim O(r_+) \,\rightarrow\, \tilde{U} \lesssim O(l),
\end{equation}
respectively, yielding
\begin{equation}
  \varDelta t_{\rm max} = \frac{2 l^2}{d r_+} \ln\frac{l}{l_{\rm P}}.
\end{equation}
Here, the fractional correction is of order $1/\ln(l/l_{\rm P})$.

The scrambling time of a large AdS black hole was calculated in Ref.~\cite{Saraswat:2020zzf}:
\begin{equation}
  v_{\rm scr} = \frac{2(d-1) l^2}{d r_+} \ln\frac{l}{l_{\rm P}}.
\label{eq:v_scr-AdS}
\end{equation}
This is related to the relevant scrambling time in boundary time as
\begin{equation}
  t_{\rm scr} = v_{\rm scr} - t_{\rm sig} = \frac{2 (d-2) l^2}{d r_+} \ln\frac{l}{l_{\rm P}},
\end{equation}
where $t_{\rm sig} = (2 l^2/d r_+)\ln(l/l_{\rm P})$ is the signal propagation time.
(These expressions are obtained by setting $r_{\rm e} \sim r_+$ in Ref.~\cite{Langhoff:2020jqa}.)
We thus find again
\begin{equation}
  \frac{\varDelta t_{\rm max}}{t_{\rm scr}} = \frac{1}{d-2}.
\label{eq:time-ratio-AdS}
\end{equation}
If we erect the effective theory at $t = t_* + t_{\rm scr}$, then the maximal proper time which the object falling at $t = t_*$ can spend in the effective theory is
\begin{equation}
  \varDelta \tau = O(l) \left(\frac{l_{\rm P}}{l}\right)^{d-2},
\label{eq:del-tau-AdS}
\end{equation}
which is indeed of $O(l_{\rm P})$ or smaller for $d \geq 3$.

\end{document}